\documentclass[12pt]{article}                     

\usepackage{graphicx}
\usepackage{epsfig}
\usepackage{amsmath}
\usepackage{mathrsfs}
\usepackage{amssymb}
\usepackage[authoryear]{natbib}
\usepackage[amsmath,thmmarks]{ntheorem}
\usepackage{algorithm,algorithmic}
\usepackage{url}
\usepackage[margin=1in]{geometry}
\usepackage{rotating}
\usepackage{setspace} 
\usepackage{stmaryrd}
\usepackage{hyperref}

\renewcommand{\P}{\mathbb{P}}

\newcommand{\incellof}[1]{_{\cellof{#1}}}
\newcommand{\cellof}[1]{{\mathcal{G}[#1]}}
\newcommand{\pdiff}[2]{\frac{\partial #1}{\partial #2}}

\newcommand{\pkg}[1]{{\sf #1}}
\newcommand{\proglang}[1]{{\sf #1}}

\newcommand{\Z}{\mathbb{Z}}

\newcommand{\N}{\mathrm{N}}

\newcommand{\rmd}{\mathrm{d}}

\newcommand{\E}{\mathbb{E}}

\newcommand{\cov}{\mathrm{cov}}

\newlength{\descwid}
\begin{document}

\title{Auxiliary Variable Markov Chain Monte Carlo for Spatial Survival and Geostatistical Models}
\author{Benjamin M. Taylor}
\maketitle

\begin{abstract}
   This article was motivated by the desire to improve Markov chain Monte Carlo methods for spatial survival models in which the locations of individuals in space are known. For a dataset comprising information on $n$ individuals, standard methods of MCMC-based inference involve computing the inverse of an $n\times n$ matrix at each iteration. However with a judicious choice of auxiliary variables on a regular grid with $m$ prediction points it will be shown how to fit an essentially equivalent model but with a substantially reduced computational cost. For a fixed output grid, the computational cost of the new method is reduced from $O(n^3)$ to $O(n)$; the cost of increasing the output grid size being $O(m\log m)$. Furthermore, the new method simultaneously solves the problem of spatial prediction of functions of the latent field, which for standard methods usually presents a further computational challenge. We apply the new method to a spatial survival dataset previously analysed in \cite{henderson2002} and show how the new method can be applied to spatial and spatiotemporal geostatistical datasets with the same computational benefits.
\end{abstract}

\section{Introduction}

This article was motivated by the following problem in spatial survival analysis. Suppose we have a spatial survival dataset with $n$ observations in which the times of censored or uncensored events have been recorded along with their location in space. The statistical model to be fitted to these data includes: covariate effects, $\beta$, through a linear-predictor $X\beta$, where $X$ is a design matrix; a specified functional form for the baseline hazard, $h_0(t;\omega)$ (where $t$ is time and $\omega$ are parameters of $h_0$); and a set of spatially-continuous and correlated Gaussian frailties, $Y$. Suppose it is desired to fit a parametric proportional hazards model to these data (noting at the outset that the ideas discussed here are applicable in other important situations, see Section \ref{sect:discussion}); the model for the baseline hazard is,
\begin{equation*}\label{eqn:hazexample}
    h(t;\psi,Y) =  \exp\{X\beta+Y\}h_0(t;\omega),
\end{equation*}
where $\psi=(\beta,\omega)$. The form of the hazard function determines both the density and survival functions (see below) and in turn these quantities feed into the likelihood given in Equation \ref{eqn:survlikelihood}. With an appropriate choice for the priors, it is therefore possible to use Markov chain Monte Carlo (MCMC) methods \citep{metropolis1953,hastings1970,MCMCiP,gamermanlopes} to draw samples from the posterior density and hence perform Bayesian inference for this class of models. The problem is that the cost of computing the likelihood is $O(n^3)$, which makes this method impractical for use with even moderately sized datasets without a lot of patience on the side of the user. The question this article addresses is: how can we make Bayesian inference via MCMC for this class of models tolerable? 

The solution to the $O(n^3)$ problem proposed in this article was inspired by considering the \emph{purpose} of the spatially correlated frailties. The purpose of the frailty term in \emph{non-spatial} survival models is to capture any unexplained variation in the outcome of interest, that is variation that cannot be explained by the available covariates. \emph{Spatially correlated} frailties are used when it is thought that the hazard of individuals close together in space (conditional on their individual covariates) is similar and we wish to take this into account when computing the likelihood. But if we suspect individuals close-by in space to be similar in some sense, do we need individual-level spatially correlated frailties? The solution proposed in this article is to replace the spatially correlated frailties in Equation (\ref{eqn:hazexample}) by a (sometimes) much larger set of auxiliary $Y$s in such a way as to actually \emph{reduce} computational cost. This is achieved by assuming that individuals who are very close together share the spatially correlated component of their frailties, but are allowed to retain individual non-spatial frailties. The sharing of the auxiliary frailties in this way is similar to specifying a hierarchical model for the data, but they are introduced in such a way so that Fourier methods \citep{wood1994} can be used for matrix operations, rendering the resulting computations $O(m\log m)$ where $m$ is the number of auxiliary frailties -- but with with the added advantage that computation of the posterior for fixed $m$ increases as $O(n)$.

The main contribution of this article is therefore a method that makes it possible to fit spatial survival models to much larger datasets than would normally be practical and with the additional benefit of providing a solution to the problem of spatial prediction (that is predicting the latent spatial field $Y$ at locations where we do not have data), which otherwise is a further computationally intensive step in the inferential process, see Section \ref{sect:spatialpredict}. Furthermore, the main idea is easily extensible to spatial and spatiotemporal geostatistical datasets as discussed in Section \ref{sect:discussion}. The new method requires the frailties to be defined on a regular grid and also that $Y$ is a stationary-Gaussian process.

It is increasingly common for survival datasets to include the spatial (and/or temporal) coordinates of observations. Example application areas of spatial survival methods in the literature have included: infant mortality in Minnesota \citep{banerjee2003}, smoking cessation in cure-rate/time-to-relapse models \citep{banerjee2004}, air pollution and mortality in Los Angeles \citep{jerrett2005}, American cancer data at the county level \citep{diva2008}, the timing of U.S. House members' position announcements on the North American Free Trade Agreement \cite{darmofal2009} and cancer mortality of Hiroshima atomic bomb survivors \citep{tonda2012}. 

The complex nature of spatial survival modelling has historically meant there have been more papers using Bayesian methods than Frequentist methods. Exceptions to this rule include the article by \cite{paik2012}, who introduce a composite likelihood method for inference and the article by \cite{li2006}, who employ spatial semiparametric estimating equations. Others exceptions include articles developing scan statistics designed for spatial survival data such as \cite{huang2007} and \cite{bhatt2014}. 

In the present article we continue the Bayesian tradition and although the ensuing discussion will neither concentrate on semiparametric proportional hazards nor on proportional odds models; we note that the main idea is easily extensible in these directions \citep{li2002,banerjee2005,diva2008}. The present article concentrates on purely spatially referenced data, rather than on spatio-temporally referenced data as discussed by \cite{banerjee2003a}; we note that the extension to this model class is also possible. As in \cite{li2002}, in the present article we will use log-Gaussian frailties and our discussion will focus on the spatial proportional hazards model as was investigated by \cite{henderson2002}, \cite{banerjee2003} and \cite{diva2008}. Another interesting, but at best tentatively related work, is \cite{zhao2011} who induce spatial correlation through a prior based on a mixture of spatially dependent Polya trees -- the method proposed in the current article is tentatively related as it will make use of a judiciously chosen prior on the random effects for spatial prediction.

Whilst the issue of handling the computational burden of large spatial survival datasets has received little attention in the literature, one example being \cite{hennerfeind2006}, see below: there has been some progress in the classical geostatistical literature. In the context of modelling spatiotemporal dependence in ocean temperatures, \cite{higdon1998} proposes a process convolution approach to modelling the random effects.  \cite{fuentes2007} avoids the expensive matrix computations involved by adopting an approximate likelihood approach in the spectral domain, a concern that has been raised about this and similar methods is the quality of the approximation \citep{banerjee2008}. \cite{banerjee2008} use predictive process models: they place knots at a selection of observation locations and also possibly some non-observation locations, replacing individual spatial random effects with interpolated values from the process on the lower-dimensional knots. Whilst they seek to use an $m$ smaller than the original $n$ to ease matrix computations, the approach in the present article does not adopt that strategy -- for example in Section \ref{sect:leukaemia} we sample 16384 random effects rather than 1043, but computation time is over 5 times faster. A further method for approximating spatial processes that has received some attention in the literature is by using low/fixed-rank approximations to the covariance matrix as in \cite{cressie2008}, \cite{wikle2010} and \cite{rodrigues2012}. Whilst low/fixed rank strategies are computationally effective they can be difficult to set up including choosing an appropriate rank and specifying priors; spatial dependence from such models is also more difficult to interpret. The technique of circulant embedding used in the present article has also been used to aid matrix computations for spatial processes in various other contexts including inference for log-Gaussian Cox processes: \cite{moller1998}, \cite{brix2001} and \cite{taylor2013}; and for Gaussian processes on a lattice with missing values: \cite{stroud2014}.

\cite{wikle2002}, \cite{royle2005} and \cite{paciorek2007} in the context of geostatistical analysis suggest a similar strategy to the method proposed in the present article, although their random effects are interpolated from a spectral representation of the process; in contrast, our approach is to handle the random effects directly using the 2-dimensional discrete Fourier transform as a tool for matrix computations. On a second point, these authors did not consider the issues that arise from wrap-around effects, which in the present article we handle by at least doubling the observation bounding box in each direction, as in Figure \ref{fig:fourier}. 

\cite{park2012} and \cite{xu2015} also use a similar strategy to that presented in the present article using the example of a simple Gaussian geostatistical model; they advocate Gaussian Markov random fields (GMRF) to define the covariance structure on their computational grid. Whilst for very small neighbourhood sizes their methods are efficient on this grid, for larger neighbourhood sizes, which are required for approximating Gaussian fields with larger spatial dependence (see for example \cite{taylor2013c}), they suggest Fourier methods to perform matrix computations. For these models with larger neighbourhood sizes, in which they resort to Fourier methods, the restriction to GMRF-induced covariance structures is unnecessary as will be shown in the present article. \cite{park2012} and \cite{xu2015} do not propose a concrete solution to deal with wrap around effects. 

Of the methods that address the problem of inference for large non-Gaussian datasets (specifically spatial survival datasets), \cite{hennerfeind2006} propose to use the reduced-rank methodology developed by \cite{kammann2003}; their methods are implemented in the BayesX package \citep{brezger2005}. As pointed out by \cite{park2012}, whilst such methods (utilising a low-dimensional approximation to the latent Gaussian process) are effective at reducing computation time, for large datasets the dimension of the approximation can still be very high, meaning these methods will not work well. As mentioned above there are other issues for low-rank methods including choosing the rank, specifying priors and interpreting spatial dependence.

The present article can therefore be seen as a generalisation of \cite{park2012} and \cite{xu2015} to spatial survival and other non-Gaussian spatial and spatiotemporal geostatistical models in which the (stationary) covariance function is able to adopt any permissible form (e.g. exponential, Mat\'ern etc.) on the grid: not merely GMRF form, which is a special case. As well as being computationally efficient, our proposed method also has the advantage of interpretability with respect to the parameters of the latent Gaussian process.

As stated above, for the purpose of this article the methodological focus will be spatial survival modelling, we refer the reader to Section \ref{sect:discussion} in which we show how the method can be extended to more general spatial and spatiotemporal geostatistical datasets. To that end, we begin in Section \ref{sect:survintro} with a brief introduction to spatial proportional hazards survival models and spatial prediction. In Section \ref{sect:auxmethods} we introduce the new model including details on computations using Fourier methods, inference via MCMC and re-visit spatial prediction. In Section \ref{sect:leukaemia}, we apply the method to the leukaemia dataset of \cite{henderson2002}. The article concludes with a discussion in Section \ref{sect:discussion}.

\section{Spatial Proportional Hazards Survival Models}
\label{sect:survintro}

In this section we give a brief overview of spatial proportional hazards survival models and discuss two methods for spatial prediction. 

Survival analysis (`event-history analysis', `duration analysis' or `reliability analysis' as it is referred to in other disciplines) is concerned with the modelling and prediction of time-to-event data: for each individual, we observe the time of an event. We do not observe the exact survival time of all individuals, rather we might only know that an individual was alive the last time they were seen by a doctor, for example; such events are termed `(right) censored'. The concept of censoring is what makes survival analysis statistically interesting. There are three types of censoring: right censoring is where the event happened after some known time, left censoring is where the event happened sometime before a known time and interval censoring is where the event happened during a known interval of time. We refer the reader to \cite{cox1984}, \cite{klein2003} or \cite{klein2013} for further details on the preliminaries of survival analysis. 

\subsection{Survival Models}

Let the occurrence time of an event of interest, $T$, be a random variable. There are three main quantities of interest in survival analysis: the density function, $f(t)$, giving the probability density function of $T$; the hazard function,
\begin{equation*}
    h(t) = \lim_{\Delta t\rightarrow 0}\left\{\frac{\P(t\leq T \leq t+\Delta t)}{\Delta t\times S(t)}\right\},
\end{equation*}
giving the instantaneous failure rate at time $t$ conditional on survival up to time $t$; and the survival function, $S(t)$, giving the probability of survival after time $t$. These three quantities are related as follows:
\begin{equation}\label{eqn:survrels}
    S(t) = 1 - \int_0^{t}f(x)\rmd x = \exp\{-H(t)\} \qquad\text{and}\qquad h(t)=\frac{f(t)}{S(t)}.
\end{equation}
Other quantities of interest are the cumulative hazard, $H(t)=\int_0^th(s)\rmd s$ and the lifetime distribution function, $F(t)=1-S(t)$. 

A (parametric) proportional hazards spatial survival model is completely defined by introducing a model for the hazard function, e.g.,
\begin{equation}\label{eqn:spatialPHmodel}
    h(t_i;\psi,Y_i) =  \exp\{X_i\beta+Y_i\}h_0(t_i;\omega),
\end{equation}
Here a subscript $i$ denotes belonging to individual $i$, $X_i$ is a vector of covariates, $\beta$ are the covariate effects, $\omega$ are parameters of the baseline hazard function (which has a tractable form, see below) and in this article $Y_i$ is the value of a spatially continuous, stationary latent Gaussian field $Y$ at the location of observation $i$. The parameters of the covariance function of the latent field $Y$ will be denoted $\eta$. We will assume that the latent field has been parametrised in such a way that $\E[\exp(Y)]=1$, which is easily achieved by setting the mean of $Y$ to be $-\sigma^2/2$, where $\sigma^2$ is the marginal variance of the field. Examples of suitable spatial covariance functions useful in epidemiological applications include the Mat\'ern and Exponential models. We require $h_0$ to be `tractable' in the sense that we must be able to evaluate it given parameters $\omega$, and further we must also be able to evaluate the baseline cumulative hazard, $H_0=\int_0^th(s)\rmd s$. Examples of suitable functional forms for the baseline hazard include the exponential, Weibull, Gompertz, Gompertz-Makeham, log-normal, gamma and F models.                                      
           
Let $\xi_i=(\beta,\omega,\eta,Y_i)$. We can derive the density function for a parametric PH spatial survival model using the right-hand expression in Equation~\ref{eqn:survrels}:
\begin{equation}\label{eqn:survdens}
    f(t;\xi_i) = \exp\{X_i\beta+Y_i\}h_0(t;\omega)\exp\left\{-\exp\{X_i\beta+Y_i\}H_0(t;\omega)\right\};
\end{equation}
The corresponding lifetime distribution is,
\begin{equation}\label{eqn:survdist}
    F(t;\xi_i) = 1 - \exp\left\{-\exp\{X_i\beta+Y_i\}H_0(t;\omega)\right\}.
\end{equation}
Let $t_i$ be the data from the $i$th individual, which in the case of an interval censored observation is represented as a vector, $t_i=(t_i^{(1)},t_i^{(2)})$. Conditional on the latent field, $Y_1,\ldots,Y_n$, we will assume that observations are independent. The likelihood in this case splits into contributing components from left, right and interval censored data as well as uncensored data:
\begin{equation}\label{eqn:survlikelihood}
    \pi(t_1,\ldots,t_n|\xi)=\hspace{-1.5em}\prod_{i\text{ uncensored}}\hspace{-1.2em}f(t_i;\xi_i)\prod_{i\text{ left
censored}}\hspace{-1.7em}F(t_i;\xi_i)\prod_{i\text{ right censored}}\hspace{-1.7em}S(t_i;\xi_i)\prod_{i\text{ interval
censored}}\hspace{-2.2em}\left[F(t_i^{(2)};\xi_i)-F(t_i^{(1)};\xi_i)\right].
\end{equation}

\subsection{Spatial Prediction}
\label{sect:spatialpredict}

Spatial prediction, the process by we predict properties of the latent field at locations where we do not have any data, can be achieved either in-line with the MCMC scheme or post-MCMC.

For the post-MCMC approach, having fitted a spatial survival model, suppose now that we wish to predict the latent field $Y$ at some additional locations $\tilde Y_1,\ldots,\tilde Y_m$, this can be achieved using the following integral,
\begin{equation*}
    \pi(\tilde Y_{1:m}|\text{data}) = \int \pi(\tilde Y_{1:m}|Y_{1:n},\eta,\text{data})\pi(Y_{1:n},\eta|\text{data})\rmd Y_{1:n}\rmd\eta.
\end{equation*}
Assuming that $\tilde Y_{1:m}$ is conditionally independent of the data given $Y_{1:n}$, we can obtain an unbiased estimate of the above as
\begin{equation}\label{eqn:predictmix}
    \pi(\tilde Y_{1:m}|\text{data})\approx\frac1n\sum_{i=1}^n\pi(\tilde Y_{1:m}|Y_{1:n}^{(i)},\eta^{(i)}),
\end{equation}
where for example $\eta^{(i)}$ is the $i$th sample of the parameter $\eta$ from the MCMC chain. Since $\pi(\tilde Y_{1:m},Y_{1:n}^{(i)}|\eta^{(i)})$ is jointly Gaussian by assumption, the conditional $\pi(\tilde Y_{1:m}|Y_{1:n}^{(i)},\eta^{(i)})$ is available in exact form, see for example \cite{rueheld2005}; the Monte Carlo approximation in Equation (\ref{eqn:predictmix}) can thus be seen as a mixture of Gaussians. Unbiased estimates of quantiles of the marginals can be obtained using
\begin{equation*}
    \int_{-\infty}^y\pi(\tilde Y_j|\text{data})\rmd\tilde Y_j \approx \frac1n\sum_{i=1}^n\int_{-\infty}^y\pi(\tilde Y_j|Y_{1:n}^{(i)},\eta^{(i)})\rmd\tilde Y_j,
\end{equation*}
there being established numerical methods for evaluating the contributing integrals on the right hand side. Whilst the above computations are tractable, in general they are computationally expensive. The cost of computing each conditional distribution, $\pi(\tilde Y_{1:m}|Y_{1:n}^{(i)},\eta^{(i)})$, is $O(n^3)$. 


The in-line method, described by \cite{diggle1998} for example, is more straightforward in principle and firstly involves running the MCMC chain to sample from $\pi(Y,\beta,\omega,\eta|\text{data})$ until convergence. At this point the parameter vector is augmented with $\tilde Y$s and a Gibbs scheme is used to sample from $\pi(\tilde Y,Y,\beta,\omega,\eta|\text{data})$ using the decomposition,
\begin{equation*}
    \pi(\tilde Y,Y,\beta,\omega,\eta|\text{data}) = \pi(\tilde Y|Y,\beta,\omega,\eta,\text{data})\pi(Y,\beta,\omega,\eta|\text{data}),
\end{equation*}
noting that under this model, conditional independence properties imply $\pi(Y,\beta,\omega,\eta|\text{data})=\pi(Y,\beta,\omega,\eta|\tilde Y,\text{data})$. The MCMC sampler already in use can continue to be used to draw samples from the density $\pi(Y,\beta,\omega,\eta|\text{data})$ and at each `retain' iteration, sampling from $\pi(\tilde Y|Y,\beta,\omega,\eta,\text{data})$ is a straightforward application of the conditional distribution of a multivariate Gaussian, see \cite[pp. 309]{diggle1998} or \cite{rueheld2005}. For increasing numbers of observations, the dominating cost of this method is $O(n^3)$.

\section{Auxiliary Variables for Spatial Survival Models}
\label{sect:auxmethods}

In this section we show how by making a slight modification to model (\ref{eqn:spatialPHmodel}), we are able to make huge computational gains, both in terms of analysing our data and for the purpose of spatial prediction. The main idea is to replace the individual spatially correlated frailties by a collection of auxiliary frailties (and possibly some additional non-spatial frailties) so that although the number of parameters in our model typically is much larger, the computational complexity is massively reduced. We describe how these computational gains are achieved using Fourier methods, discuss a generic gradient-based adaptive MCMC algorithm for efficient inference and re-visit spatial prediction for the new method.

The most common use of auxiliary variable methods \citep{edwards1988,besag1993,higdon1998,damien1999,pitt2001} in Bayesian sampling schemes such as MCMC is in the situation where it is desired to sample from the density of a random variable $\theta$ conditional on some data, $\pi(\theta|\text{data})$, but direct sampling from this density is difficult. If it is possible to introduce an additional set of variables $\vartheta$ in such a way that sampling from $\pi(\theta,\vartheta|\text{data})$ is easier, then $\vartheta$ are known as auxiliary variables. Auxiliary variables are a very important technique in Bayesian sampling schemes and have been applied in many areas including binary/multinomial regression \citep{holmes2006}, capture/recapture modelling \citep{pollock2002} survival analysis/multiple imputation \citep{faucett2002}, stationary time series models \citep{pitt2005}, variable selection \citep{nott2004}, mixture sampling \citep{fruhwirth2007} and sampling jump processes \citep{rao2013} among many others. In the present article we refer to auxiliary variables in the same spirit as \cite{pitt2001}, in that they are additional variables `present to aid the task of simulation'.

\subsection{The Model}
\label{sect:newmodel}

Our frailties will be defined as piecewise constant on a fine, regular $2^{m_1}\times2^{m_2}$ (where $m_1,m_2\in\Z_{\geq 1}$) grid of cells, $\mathcal{G}$, that cover an extended version of the bounding box of all observations in space (see Section \ref{sect:fourier} for further details). The value of the stationary Gaussian field $Y$ at the centroid of each of these cells will be denoted $Y_1,\ldots Y_m$ where $m=2^{m_1+m_2}$.

The basic spatial survival model we will use is very similar to that in Equation \ref{eqn:spatialPHmodel}:
\begin{equation}\label{eqn:newmodel}
    h(t_i;\psi,Y\incellof{i}) =  \exp\{X_i\beta+Y\incellof{i}\}h_0(t_i;\omega),
\end{equation}
where all terms are as before, but where $\cellof{i}\in\{1,\ldots,m\}$ denotes the index of the grid cell to which observation $i$ belongs. A further extension of this model might allow
\begin{equation}\label{eqn:newmodelUi}
    h(t_i;\psi,Y_\cellof{i},U_i) =  \exp\{X_i\beta+Y_\cellof{i}+U_i\}h_0(t_i;\omega),
\end{equation}
where the $U_i\sim\N(0,\sigma_U^2)$ are individual-specific non-spatial frailties. The idea of including both spatial and non-spatial random effects is common in the geostatistics literature, see for example \cite{diggle2002} and \cite{diggle2007}, and was used in spatial survival models in \cite{darmofal2009}.

During the MCMC scheme, we will not sample the $Y_i$s directly, rather we work with a vector of transformed variables, $\Gamma=(\Gamma_1,\ldots,\Gamma_m)^T$, such that
\begin{equation}\label{eqn:gammaeq}
    \left(\begin{array}{c}Y_1\\ \vdots \\ Y_m\end{array}\right) = -\sigma^2/2 + \Sigma^{1/2}_\eta\left(\begin{array}{c}\Gamma_1\\ \vdots \\ \Gamma_m\end{array}\right),
\end{equation}
where $\Sigma^{1/2}_\eta$ is the Cholesky decomposition of the covariance matrix, $\Sigma_\eta$, obtained by evaluating the covariance function at each of the centroids of the grid cells in $\mathcal{G}$; note the dependence on $\eta$. This is really the key ingredient to the new method, since there is a very sensible prior for $\Gamma$, namely the multivariate $\N(0,1)$ prior. The reason this is such a good prior is that if we wanted to simulate a multivariate Gaussian variable with mean $-\sigma^2/2$ and variance matrix $\Sigma_\eta$ then this can be achieved precisely by first simulating $\Gamma\sim\N(0,1)$ and then using Equation (\ref{eqn:gammaeq}) to instantiate an appropriate $\{Y_1,\ldots,Y_m\}$.

\subsection{Computation Using Fourier Methods}
\label{sect:fourier}

In this section we discuss the use of Fourier methods and how these can lead to a substantial computational saving despite a potentially large increase in the dimensionality of the problem.

\begin{figure}[H]
    \centering
        \includegraphics[width=0.5\textwidth]{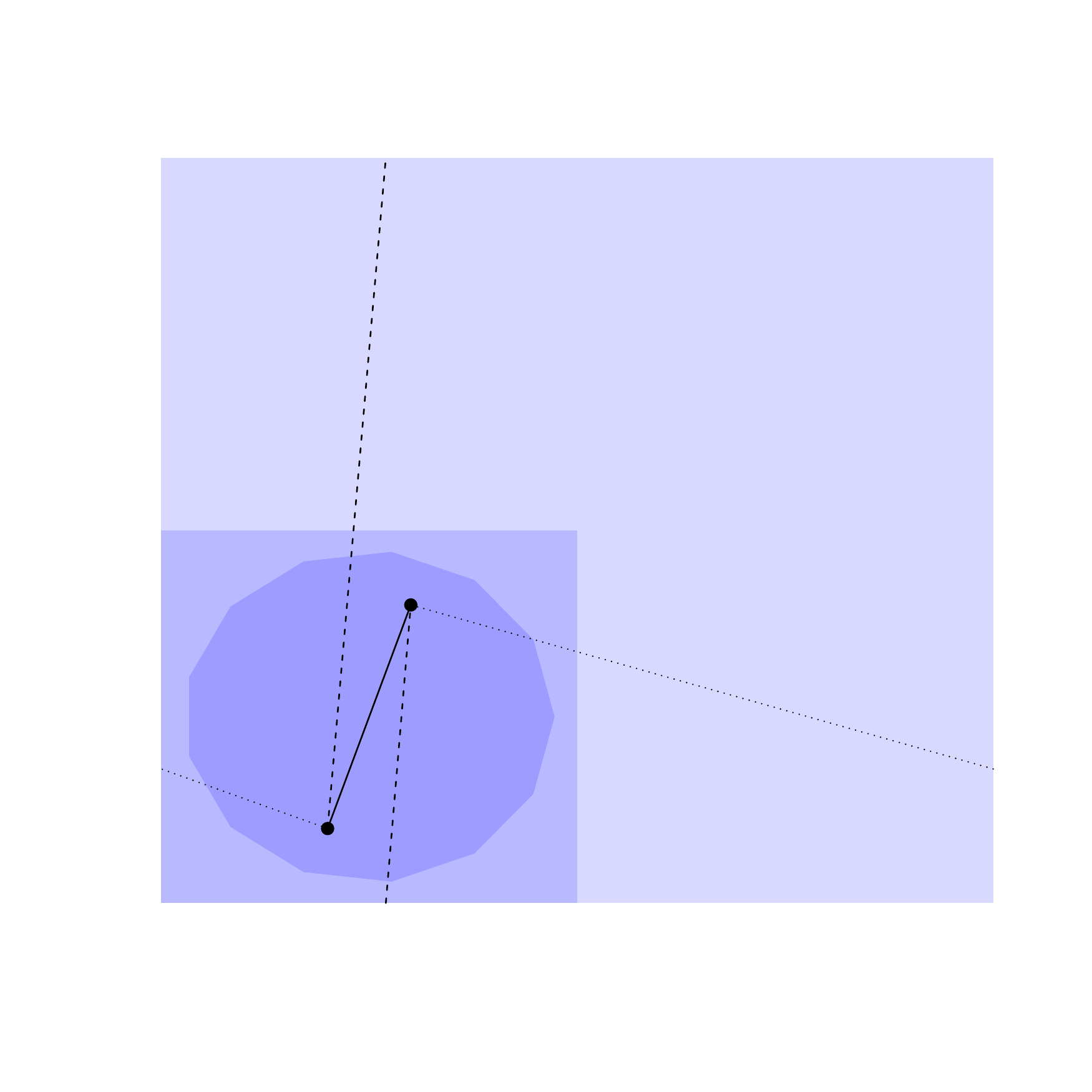}
    \caption{\label{fig:fourier} Diagram illustrating how the observation window (the triskaidecagon in the bottom left corner) is initially embedded in a bounding box (the dark square) and then inside the extended bounding box (the light square). The solid, dashed and dotted lines show three different routes on the notional torus between two points in the observation window; the shortest of these routes is used in computing the covariance between any two points in the extended bounding box.}
\end{figure}

In Section \ref{sect:newmodel}, it was mentioned that the bounding box is extended, the sense in which this is achieved is illustrated in Figure \ref{fig:fourier}. In this diagram, the polygon (a triskaidecagon) in the bottom left represents what will be referred to as the `observation window', inside of which all of the observations are located; technically, this observation window is not actually required: the purpose of it is to illustrate a bounded region on the plane within which the data are contained (note the choice of this polygon is arbitrary and the region neither needs to be convex nor even connected). Surrounding the observation window, there is a dark bounding box containing the observation window (and by extension, the observations) and a lighter extended bounding box, containing the bounding box. The location of the bounding box within the extended bounding box is arbitrary: it could equally well have been in the centre, for example.

The process $Y$ will be represented on a fine regular grid $Y_1,\ldots Y_m$, partitioning the extended bounding box into $m=2^{m_1+m_2}$ rectangles, on which $Y$ will be treated as piecewise constant. This `extended grid' is notionally wrapped on a torus and we use a toroidal distance metric, again illustrated in Figure \ref{fig:fourier}, to compute the distance between the centroids of any two rectangular cells on the extended grid. The toroidal metric is the minimum distance between two points, either travelling directly or around the minor and/or major radii. A precise definition is given in \cite{moller1998}. We use the toroidal distances in specifying the covariance matrix, $\Sigma_\eta$, on the extended grid. The restriction to $2^{m_1}\times2^{m_2}$ grids is the optimal choice for computational purposes, but this is flexible. For example, there exist algorithms for the construction of computational `plans' for fast implementation of the discrete Fourier transform (DFT) on other grid sizes, see the FFTW library \citep{frigo2011} and \proglang{R} wrapper library \citep{krey2011}. In working on the extended grid, rather than on a grid over the bounding box, we effectively eliminate wrap-around effects in the observation window since the distance between any two points inside it is the same as the planar Euclidean distance -- for such pairs of points it is always shorter to travel directly and not around the torus. In other words the covariance between two points in the observation window, computed using the toroidal distance metric, will be correct.

A full discussion of how the DFT is used in matrix computations on block circulant matrices is given in Chapter 2 of \cite{rueheld2005} or \cite{taylor2013c}, but very briefly the idea is to use the spectral decomposition of $\Sigma_\eta$ into the product 
\begin{equation*}
\Sigma_\eta=FEF^H, 
\end{equation*}
where the superscript $H$ denotes the conjugate transpose. Matrix-vector products such as $Fv$ and $F^Hv$ for some vector $v$ are available as a suitably normalised discrete Fourier transform (or respectively inverse discrete Fourier transform) of $v$.

On the extended grid, the matrix $\Sigma_\eta$ will have block circulant structure, and it is this property that allows us to make use of the DFT for computations such has computing the matrix square root, or inversion. In brief, all of the information about the second order structure of the process $Y$ on the extended grid is contained in a $2^{m_1}\times2^{m_2}$ matrix called a `base matrix', $\varsigma_\eta$, say. As to why Fourier methods are so useful: for example, the two-dimensional DFT of $\varsigma_\eta$ gives a second matrix of the same size containing the eigenvalues of $\Sigma_\eta$, computed in $O(m\log m)$ time. These base matrices are not to be manipulated in the ordinary sense: they represent a massive condensing of information, since on a $2^{m_1}\times2^{m_2}$ grid the full covariance matrix would have $(2^{m_1+m_2})^2$ entries.

The main disadvantage of Fourier methods is that large values of the spatial decay parameter, compared with the size of the observation window can lead to a non-positive definite covariance matrix for the frailties. One solution to this issue is to extend the grid further, for example by tripling or quadrupling the bounding box in each direction, with an associated increase in computational cost. However for the usual doubling of the bounding box in each direction, the occurrence of non-positive definite matrices in an MCMC run can be prevented by using an informative prior for the spatial decay parameter that does not place much weight outside of roughly 1/5 of the width of the observation window. It is the opinion of the present author that this is only a very minor concern, for to obtain reliable estimates of the spatial decay parameter, we require replication of the latent process $Y$ and therefore an observation window large enough to permit this to occur.

\subsection{Inference Via MCMC}
\label{sect:MCMC}

In this section we discuss MCMC for model (\ref{eqn:newmodel}). Our Bayesian model for spatial survival data includes the chosen hazard form, the likelihood from Equation \ref{eqn:survlikelihood} and a set of priors for the model parameters. We do not in fact typically sample $\omega$ and $\eta$ directly, rather because these parameters often only have support on the positive real line, we perform MCMC on a transformed scale, for example appropriate transforms might be $\tilde\omega=\log\omega$ and $\tilde\eta=\log\eta$. By Bayes' Theorem, the product of the prior and likelihood is proportional to the posterior:
\begin{equation*}
   \pi(\beta,\tilde\omega,\tilde\eta,\Gamma|\text{data})=\frac{\pi(\text{data}|\beta,\tilde\omega,\tilde\eta,\Gamma)\pi(\beta,\tilde\omega,\tilde\eta,\Gamma)}{\pi(\text{data})}\propto\pi(\text{data}|\beta,\tilde\omega,\tilde\eta,\Gamma)\pi(\beta,\tilde\omega,\tilde\eta,\Gamma);
\end{equation*}
the quantity $\pi(\text{data})$ is the marginal likelihood. Note that the conditional independence properties of this model imply that $\pi(\text{data}|\beta,\tilde\omega,\tilde\eta,\Gamma)=\pi(\text{data}|\beta,\tilde\omega,\Gamma)$.

We suggest using Markov chain Monte Carlo methods \citep{MCMCiP,gamermanlopes} to draw from the posterior, specifically advanced MCMC schemes that make use of gradient information to inform the proposal kernel. The scheme that we consider here is a mix of adaptive random walk and Metropolis-adjusted Langevin kernels: Langevin kernels for $\beta$, $\tilde\omega$ and $\Gamma$ and a random walk kernel for $\tilde\eta$. The reasons we advocate a random walk kernel for $\tilde\eta$ are (i) because the gradient with respect to this parameter can be difficult to evaluate for general covariance functions and there is a non-negligible cost associated with computing it, which would be required for a Langevin proposal; and (ii) the parameters $\tilde\eta$ (in particular the spatial decay parameter) are typically not very well identified by the data in any case \citep{zhang2004} -- so it is not clear the extra effort of computing the gradient with respct to $\tilde\eta$ is worthwhile. There are of course other choices for proposal kernel for example Hamiltonian schemes \citep{neal2011,girolami2011}, which should be more efficient if tuned correctly, but one main advantage of random walk and Langevin kernels is that there are theoretical results \citep{roberts2001} that can be used to guide the tuning process, which otherwise can be both difficult and tiresome.

The MCMC method we use is an example of a Metropolis-Hastings sampling scheme in which having initialised the chain at, $\{\beta^{(0)},\tilde\omega^{(0)},\tilde\eta^{(0)},\Gamma^{(0)}\}$, the $i$th step of the algorithm involves drawing a candidate $\{\beta^*,\tilde\omega^*,\tilde\eta^*,\Gamma^*\}$ from a proposal density, $q$, and accepting it, i.e., setting $\{\beta^{(i)},\tilde\omega^{(i)},\tilde\eta^{(i)},Y^{(i)}\}=\{\beta^*,\tilde\omega^*,\tilde\eta^*,\Gamma^*\}$, with probability

\begin{equation*}
   \min\left\{1,\frac{\pi(\beta^*,\tilde\omega^*,\tilde\eta^*,\Gamma^*|\text{data})}{\pi(\beta^{(i-1)},\tilde\omega^{(i-1)},\tilde\eta^{(i-1)},\Gamma^{(i-1)}|\text{data})}\frac{q(\beta^{(i-1)},\tilde\omega^{(i-1)},\tilde\eta^{(i-1)},\Gamma^{(i-1)}|\beta^*,\tilde\omega^*,\tilde\eta^*,\Gamma^*)}{q(\beta^*,\tilde\omega^*,\tilde\eta^*,\Gamma^*|\beta^{(i-1)},\tilde\omega^{(i-1)},\tilde\eta^{(i-1)},\Gamma^{(i-1)})}\right\}.
\end{equation*}

Let $\zeta=(\beta,\tilde\omega,\tilde\eta,\Gamma)$. We suggest the following proposal:
\begin{equation}
   q(\zeta^{(i^*)}|\zeta^{(i-1)})=\N\left[\zeta^{(i^*)};\mu_{\zeta^{(i-1)}},h^2\Sigma\right].
\label{eqn:overall}
\end{equation}
where
\begin{equation}\label{eqn:propdetails}
   \mu_{\zeta^{(i-1)}} = \left(\begin{array}{c}
   (\beta,\tilde\omega)^{(i-1)}+\frac{h^2h_{\beta,\tilde\omega}^2}{2}\Sigma_{\beta,\tilde\omega}\pdiff{\log\{\pi(\zeta^{(i-1)}|Y)\}}{(\beta,\tilde\omega)}\\
   \tilde\eta^{(i-1)}\\
   \Gamma^{(i-1)}+\frac{h^2h_{\Gamma}^2}{2}\Sigma_\Gamma\pdiff{\log\{\pi(\zeta^{(i-1)}|Y)\}}{\Gamma}\end{array}\right) \ \text{and}\  \Sigma = \left(\begin{array}{ccc}   
   h_{\beta,\tilde\omega}^2\Sigma_{\beta,\tilde\omega} & 0 & 0 \\ 
   0 & ch_{\tilde\eta}^2\Sigma_{\tilde\eta} \\
   0 & 0 & h_{\Gamma}^2\Sigma_\Gamma \\ 
   \end{array}\right)
\end{equation}
In Equation (\ref{eqn:propdetails}), $\Sigma_{\beta,\tilde\omega}$ is an approximation to the negative inverse of the observed information matrix, $\{-\E[\mathcal I(\hat\beta,\hat{\tilde\omega})]\}^{-1}$ conditional on an estimated $\Gamma$ and $\tilde\eta$, and similarly for $\Sigma_{\tilde\eta}$ and $\Sigma_\Gamma$. Due to the size of the latter, we advocate using a diagonal matrix rather than the full covariance matrix. The constants $h_{\beta,\tilde\omega}^2$, $h_{\tilde\eta}^2$ and $h_{\gamma}^2$ are approximately optimal scalings for Gaussian targets explored by Gaussian random walk or MALA proposals, see \cite{roberts2001}. We set $h_{\beta,\tilde\omega}^2=1.65^2/[\dim(\beta)+\dim(\tilde\omega)]^{1/3}$, $h_{\tilde\eta}^2=2.38^2/\dim(\tilde\eta)$ and $h_{\Gamma}^2=1.65^2/\dim(\Gamma)^{1/3}$, where $\dim$ is the dimension. The constant $c$ is present because we wish to adapt the value of $h$ in our chain using a method of \cite{andrieu2008} (see \cite{taylor2013} for details) to achieve an approximately optimal acceptance rate of $0.574$ to suit the Langevin kernels, which is too high for the random walk; we set $c=0.4$ roughly the ratio of the approximate optimal acceptance rates for random walk and Langevin kernels ($0.234/0.574$).

We can use a combination of maximum likelihood estimation and \emph{ad hoc} methods to obtain initial estimates of $\beta$, $\tilde\omega$, $\tilde\eta$ and $\Gamma$. Initial estimates of the parameters $\beta$ and $\tilde\omega$ can be obtained via maximum likelihood, ignoring any spatial correlation between observations. Having obtained initial estimates of $\beta$ and $\tilde\omega$, for each individual observation we can compute a value of $Y$ which would maximise the individual contribution to the log-likelihood (conditional on $\beta$ and $\tilde\omega$), taking the cell-wise mean to get an initial \emph{ad hoc} value of $Y$ for each cell with data. In those cells for which there are no associated data the initial $Y$ can be set to the overall mean of the $Y$s from cells with data. Lastly, with the estimates thus obtained for $\beta$, $\tilde\omega$ and $Y$ (and hence $\Gamma$, given some candidate values of $\tilde\eta$), we can construct a quadratic approximation to the conditional posterior $\pi(\tilde\eta|\beta,\tilde\omega,\Gamma,\text{data})$ on a grid spread out over the range of the prior for $\tilde\eta$; this approximation can be maximised exactly to get an initial value for $\tilde\eta$, and $\Sigma_{\tilde\eta}$ can be derived exactly from the curvature of the quadratic surface at the maximum. 

Conditional on the initial guess at $\beta$, $\tilde\omega$, $\tilde\eta$ and $\Gamma$, the matrices $\Sigma_{\beta,\tilde\omega}$ and $\Sigma_\Gamma$ can be obtained from the second derivative of the posterior with respect to those parameters. Although we have an initial guess at $\Gamma$ from the above, we start the MCMC algorithm with $\Gamma=0$, which works well. 

An open-source piece of software is available for fitting this model using the above method in the form of the \proglang{R} package \pkg{spatsurv}, see \cite{taylor2014a}.

\subsection{Spatial Prediction}

Having fitted this model using MCMC, we will have a sample $\{Y_1^{(i)},\ldots,Y_m^{(i)}\}_{i=1}^N$, where $N$ is the number of retained iterations, which can be used for spatial prediction \textbf{directly}. The contribution to the likelihood of grid cells without observations is slightly subtle, whilst they obviously contribute to the posterior, their joint posterior probability density $\pi(Y_{1:m}|\text{data})$, is partly mediated through the prior $\pi(\Gamma)$ and partly through their involvement in the transformation in Equation (\ref{eqn:gammaeq}). The upshot is that we obtain unbiased joint Bayesian inference for all cells on the grid.

\section{Simulation Study: How Much Faster is the New Method?}
\label{sect:simulationstudies}

In this section we present some simulation results to give the reader an idea of the comparative computational burden of the proposed method using Fourier methods for matrix operations with the standard method using ordinary matrix operations. The discussion on complexity here refers to the cost of implementing one iteration of the MCMC sampler. All computations in this article were performed on a 3.40GHz Intel\textregistered Core\texttrademark i7-4770 desktop computer.

As mentioned briefly earlier in this article, the standard method using ordinary matrix operations has complexity $O(n^3)$, where $n$ is the number of iterations. For the proposed Fourier methods the complexity is a little more subtle: at each iteration the matrix operations (such as inversion and computation of the square root) will have a cost scaling with $m\log m$ but evaluation of the remainder of the posterior scales as $O(n)$. Thus for a fixed grid size, the cost of adding additional observations scales linearly with the number of observations.

We used MCMC to fit model (\ref{eqn:newmodel}) on datasets containing 50, 100, 200, 300, 500, 1000 and 2000 observations and recorded the time in seconds the MCMC algorithm taks to run for 1000 iterations; the time recorded does not include pre-processing. We did not run a similar analysis for Model (\ref{eqn:newmodelUi}) as it has the same computational complexity as model (\ref{eqn:newmodel}), so the plots would look similar. This simulation study only considers how long it takes to generate a sample of size 1000: for the standard model, if it is desired to perform spatial prediction, there is an additional cost on top of this -- the output of the new model includes the joint posterior of all $Y$s on the extended grid (though only the $Y$s covering the observation window are likely to be of relevance).

\begin{figure}[H]
    \centering
        \includegraphics[width=0.5\textwidth]{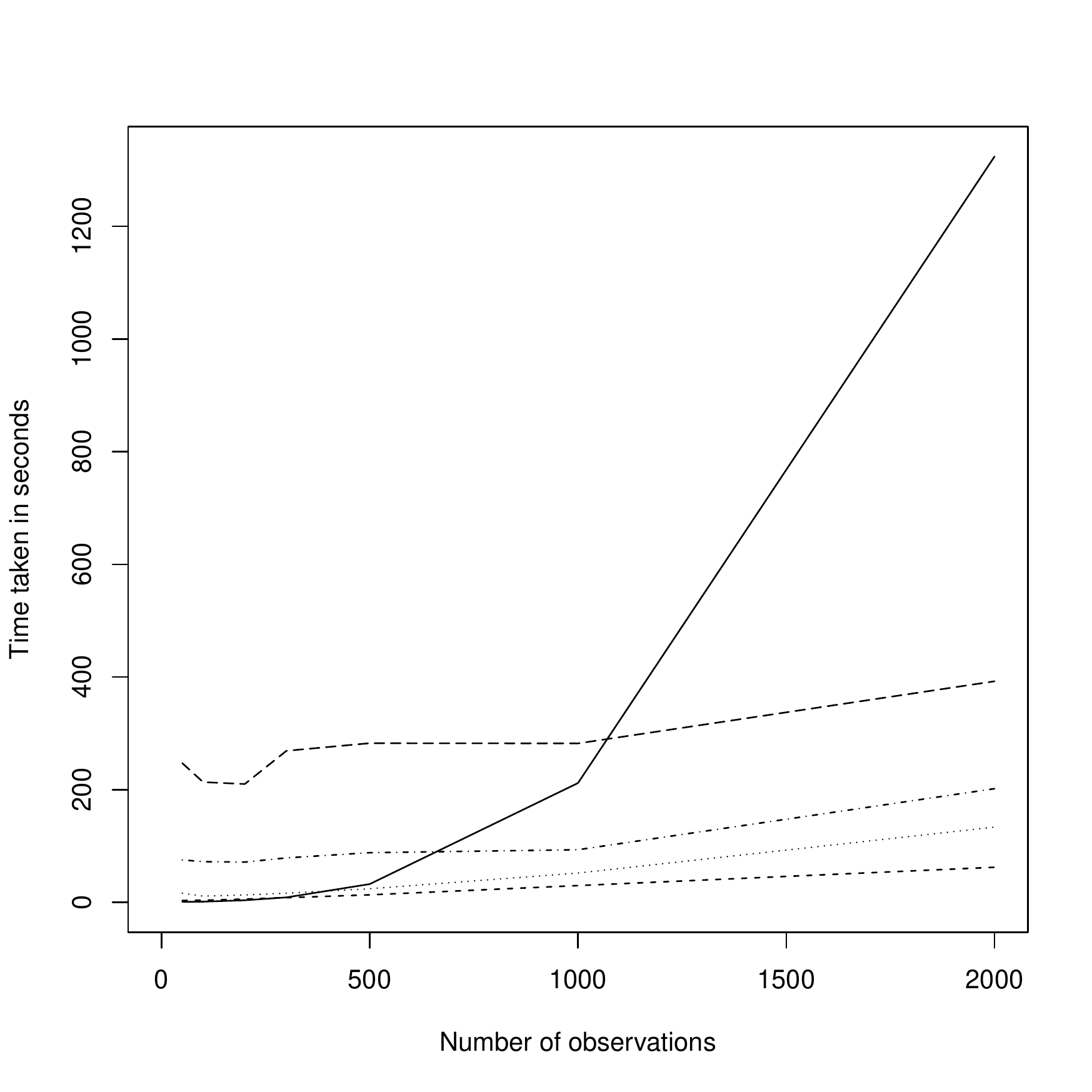}
    \caption{\label{fig:timecomparison} Plot showing the time taken to run 1000 iterations of the MCMC algorithm using: standard matrix operations (solid line) and Fourier methods on a $32\times32$ grid (short dashed), $64\times64$ grid(dotted), $128\times128$ grid (dot-dash) and $256\times256$ grid (long dash). Note for example that the output $256\times256$ grid actually runs on a $512\times512$ extended grid (i.e. is predicting over 1/4 million variables); each of the other grids is extended by 2 in each direction.}
\end{figure}

Figure \ref{fig:timecomparison} shows the results from this experiment; the plot confirms the rate of increase in computational cost for both methods. For small datasets, the new method is slower than the standard method. however, by the time there are 500 observations in the dataset, depending on the size of the output grid, the new method is faster: this is the case for output grids of size $32\times32$ and $64\times64$, but not for $128\times128$ or $256\times265$. By the time the dataset is of size 1000, the new method is at least twice as fast for grid sizes up to $128\times128$; for datasets of size 2000, the new method is 21.2 times faster for a $32\times32$ grid,  9.9 times faster for a $64\times64$ grid,  6.6 times faster for a $128\times128$ grid and 3.4 times faster for a $256\times256$ grid. 

\section{Application: Spatial Survival Analysis of Leukaemia Data}
\label{sect:leukaemia}

\begin{figure}[htbp]
    \centering
        \includegraphics[width=0.7\textwidth]{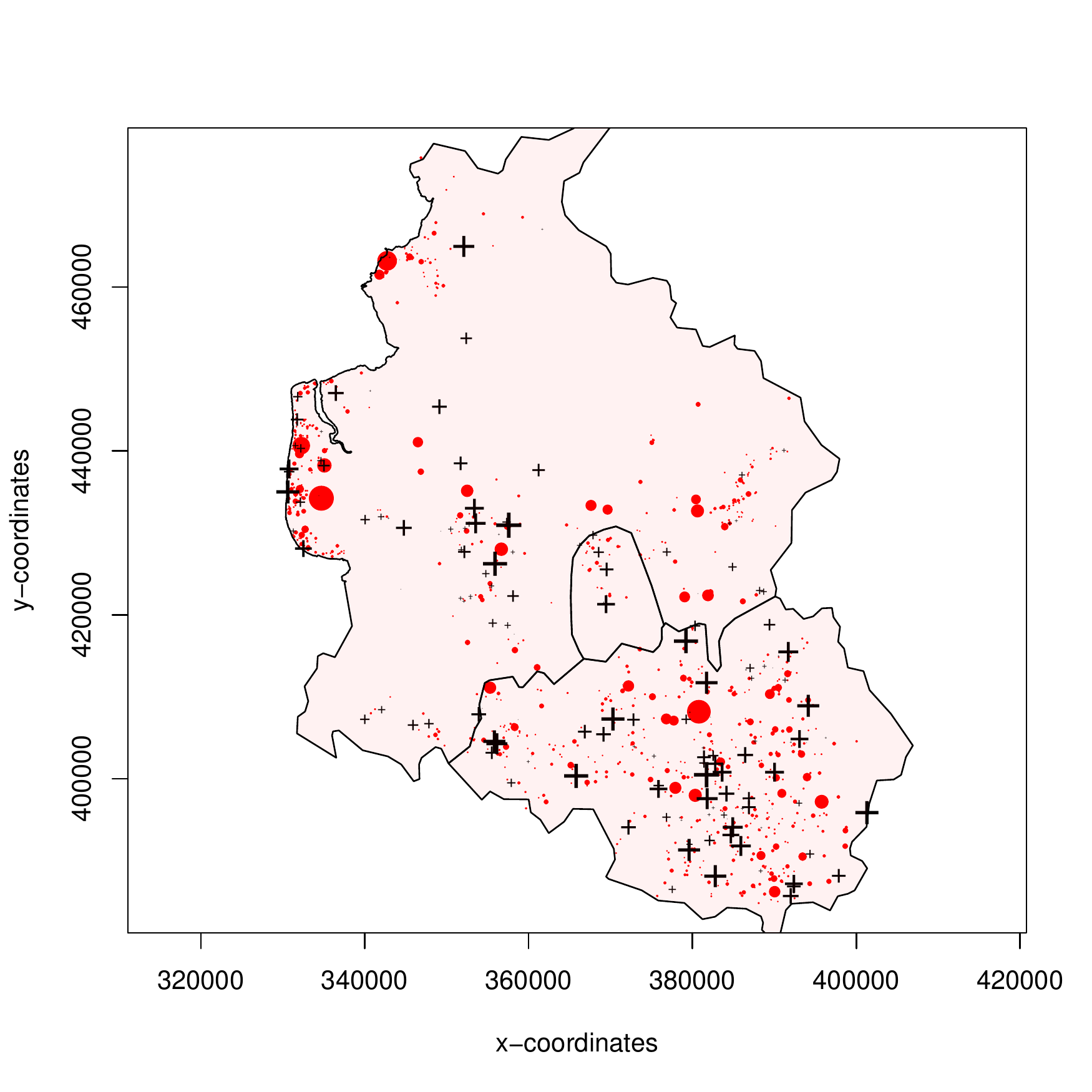}
    \caption{\label{fig:leukdata} Plot of the leukaemia data, spanning north west and south east Lancashire in the United Kingdom. The red dots represent uncensored observations, the black crosses are right-censored observations; the size of the plotting character is proportional to the observed time.}
\end{figure}

In this section, we analyse the leukaemia data from \cite{henderson2002} to demonstrate that the new method is both practical and gives comparable estimates to an established method; the reader should refer to this article for a full description of these data. The leukaemia data are illustrated in Figure \ref{fig:leukdata}, which shows the 1043 observations plotted over the north west and south east areas of Lancashire in the United Kingdom. These data were initially provided on the unit square and for the purpose of interpretability of the spatial decay parameter, were approximately transformed onto the Ordnance Survey of Great Britain projection, EPSG 27700.

Alongside the 1043 observed survival times and (right) censoring indicator, the dataset includes covariate information on each subject's age, sex, white blood cell count (WBC) and a measure of deprivation (the Townsend index). In their original article, \cite{henderson2002} used gamma frailties, the Breslow estimator \citep{breslow1974} for the baseline hazard and the Mat\'ern covariance function with smoothness parameter 1 for the covariance function of the spatially correlated frailties. In comparison, in the present article we use model (\ref{eqn:newmodel}) with log-Gaussian frailties and a Weibull parametric model for the baseline hazard. The Weibull baseline hazard takes the form,
\begin{equation}
    h_0(t;\alpha,\lambda) = \alpha\lambda t^{\alpha-1};
\end{equation}
this choice gives some flexibility regarding the shape of $h_0$ and allows us to produce posterior estimates of the baseline hazard and individual hazard/survival/density curves with credible intervals. For even greater flexibility in modelling $h_0$ it is possible to use $B$-splines \cite{deboor1978}, which are also implemented in the \pkg{spatsurv} package, whilst the increased flexibility offered by these models can sometimes be desirable they can also lead to over-fitting. 

We assumed an exponential covariance function for $Y$, i.e.
\begin{equation*}
    \cov(Y_i,Y_j) = \sigma^2\exp\{||c_i-c_j||/\phi\}
\end{equation*}
where for example $c_i$ is the coordinates of the $i$th observation and $||\cdot||$ denotes Euclidean distance. The quantity $\sigma^2$ is the marginal variance of the latent field and $\phi$ is the spatial decay parameter, larger values meaning more spatial dependence. The choice to use the exponential model was arbitrary -- we could equally have chosen the Mat\'ern model with roughness parameter 1 used in \cite{henderson2002}, but since the two models are fundamentally different in any case it was thought to be of interest to compare the spatial predictions from both models.

We placed the 1043 observations inside a $64\times64$ grid of square cells of dimension $1650\times1650$ metres and inference took place on an extended $128\times128$ grid. We used a $\N(0,10^2)$ prior for both $\beta$ and $\log\omega$, a $\N(0,0.5^2)$ prior for $\log\sigma$ and a $\N(\log5000,0.3^2)$ prior for $\phi$. The MCMC sampler was run for 1100000 iterations with a 100000 iteration burn-in and retaining every 1000th sample, the run took 11.5 hours -- see Section \ref{sect:discussion} for a method of speeding this up on a multi-core machine. The parameters $\beta$, $\log\omega$, $\log\eta$ and $\Gamma$ were initialised as described in section \ref{sect:MCMC}.

To check for convergence, we examined a plot of the value of the log-posterior over the retained iterations, this showed that the sampler had appeared to have left the transient phase and that it had found a mode, see Figure \ref{fig:logpost}. We furthermore examined trace and autocorrelation plots of the parameters $\beta$, $\omega=(\alpha,\lambda)$ and $\eta$ and a selection of 20 randomly chosen $Y$s, these can be viewed at \url{http://www.lancaster.ac.uk/staff/taylorb1/amlmcmc/MCMCoutput.html}. For the remaining $Y$s, we examined a plot of the lag-1 autocorrelations, these ranged between -0.12 and 0.13 with the 95\% quantiles between $\pm0.07$. We also overlaid plots of the prior and posterior to check whether each of the parameters was well-identified by the data; this was found to be the case with all but the spatial decay parameter $\phi$, as expected -- see Figure \ref{fig:priorposterior}.

Having provided evidence for convergence of the MCMC algorithm, we can proceed to make statistical inferences. Estimates of the the parameters $\beta$, $\omega=(\alpha,\beta)$ and $\eta$ from this model are given in Table\ref{tab:leukres}. These results can be compared with Table 5 in \cite{henderson2002} and it can be seen that these show a similar story with some minor differences in the estimates of $\beta$ and their credible intervals, as might be expected from these two different models. The other parameters ($\omega$ and $\eta$) are not directly comparable.

\begin{table}[htbp]
    \centering
    \begin{tabular}{c|ccc}
        parameter & median & 2.5\% & 97.5\% \\ \hline\hline
        age & 3.38$\times10^{-2}$ & 2.94$\times10^{-2}$ & 3.82$\times10^{-2}$ \\
        sex & 6.45$\times10^{-2}$ & -8.29$\times10^{-2}$ & 0.194 \\
        WBC & 3.2$\times10^{-3}$ & 2.31$\times10^{-3}$ & 4.13$\times10^{-3}$ \\
        deprivation & 2.92$\times10^{-2}$ & 8.25$\times10^{-3}$ & 5.16$\times10^{-2}$ \\ \hline
        $\alpha$ & 0.611 & 0.578 & 0.649 \\
        $\lambda$ & 3.02$\times10^{-3}$ & 1.95$\times10^{-3}$ & 4.5$\times10^{-3}$ \\ \hline
        $\sigma$ & 0.387 & 0.266 & 0.546 \\
        $\phi$ (metres) & 5316 & 2958 & 9521 \\ \hline
    \end{tabular}
    \caption{\label{tab:leukres} Table showing parameter estimates for the leukaemia analysis.}
\end{table}

\begin{figure}[htbp]
    \centering
    \begin{minipage}{0.5\textwidth}
        \includegraphics[width=0.9\textwidth]{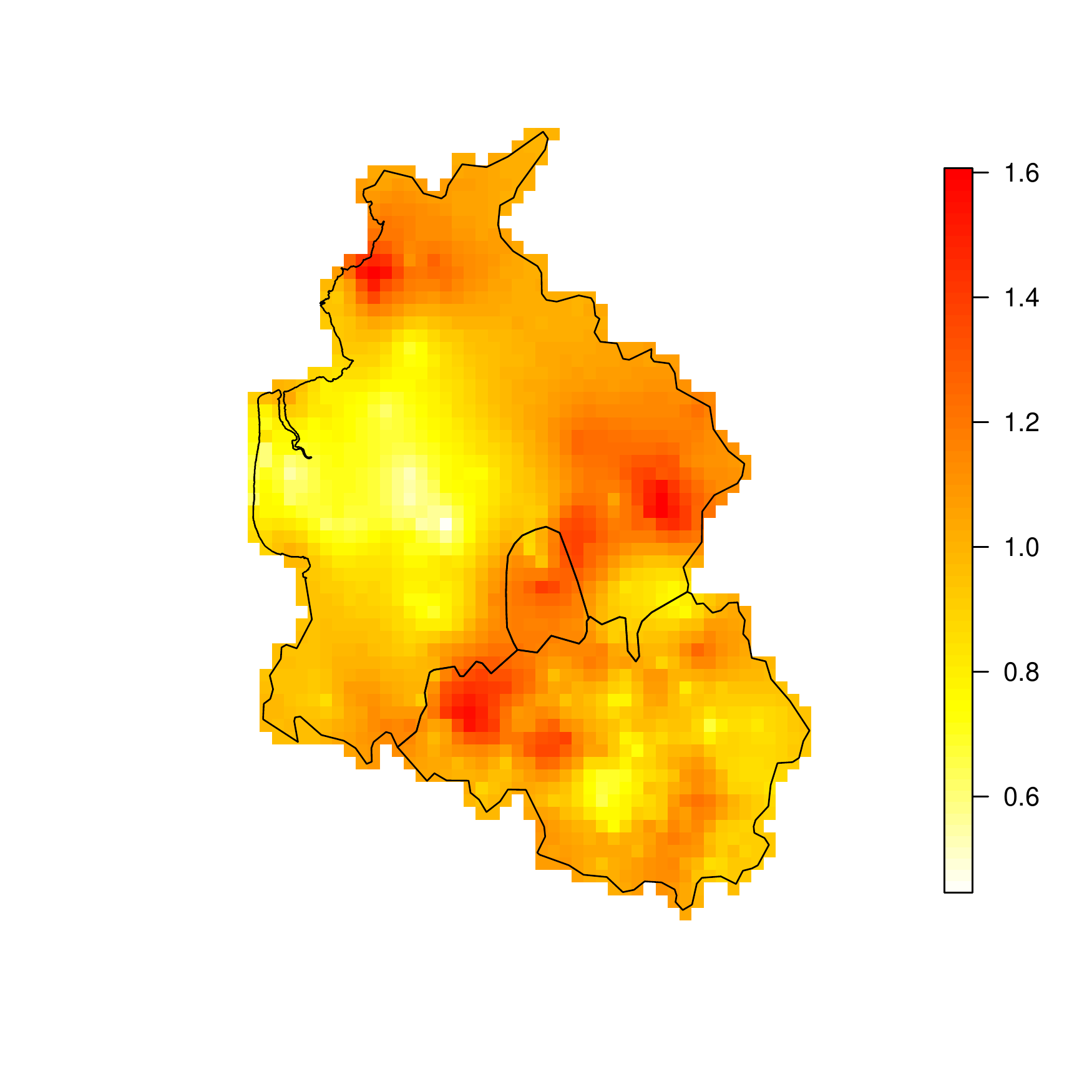}
    \end{minipage}\begin{minipage}{0.5\textwidth}
        \includegraphics[width=0.9\textwidth]{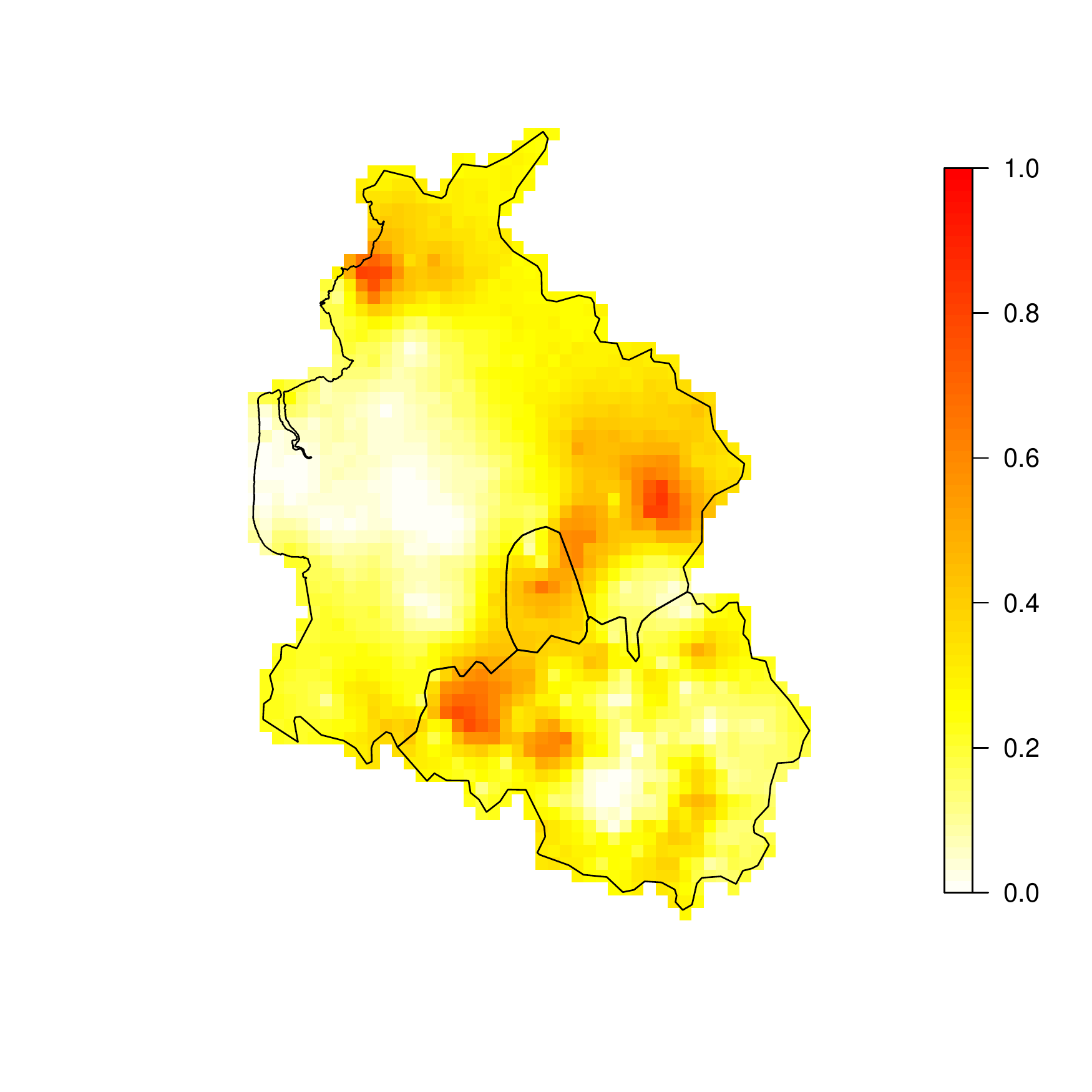}
    \end{minipage}
    
    \begin{minipage}{0.5\textwidth}
        \includegraphics[width=0.9\textwidth]{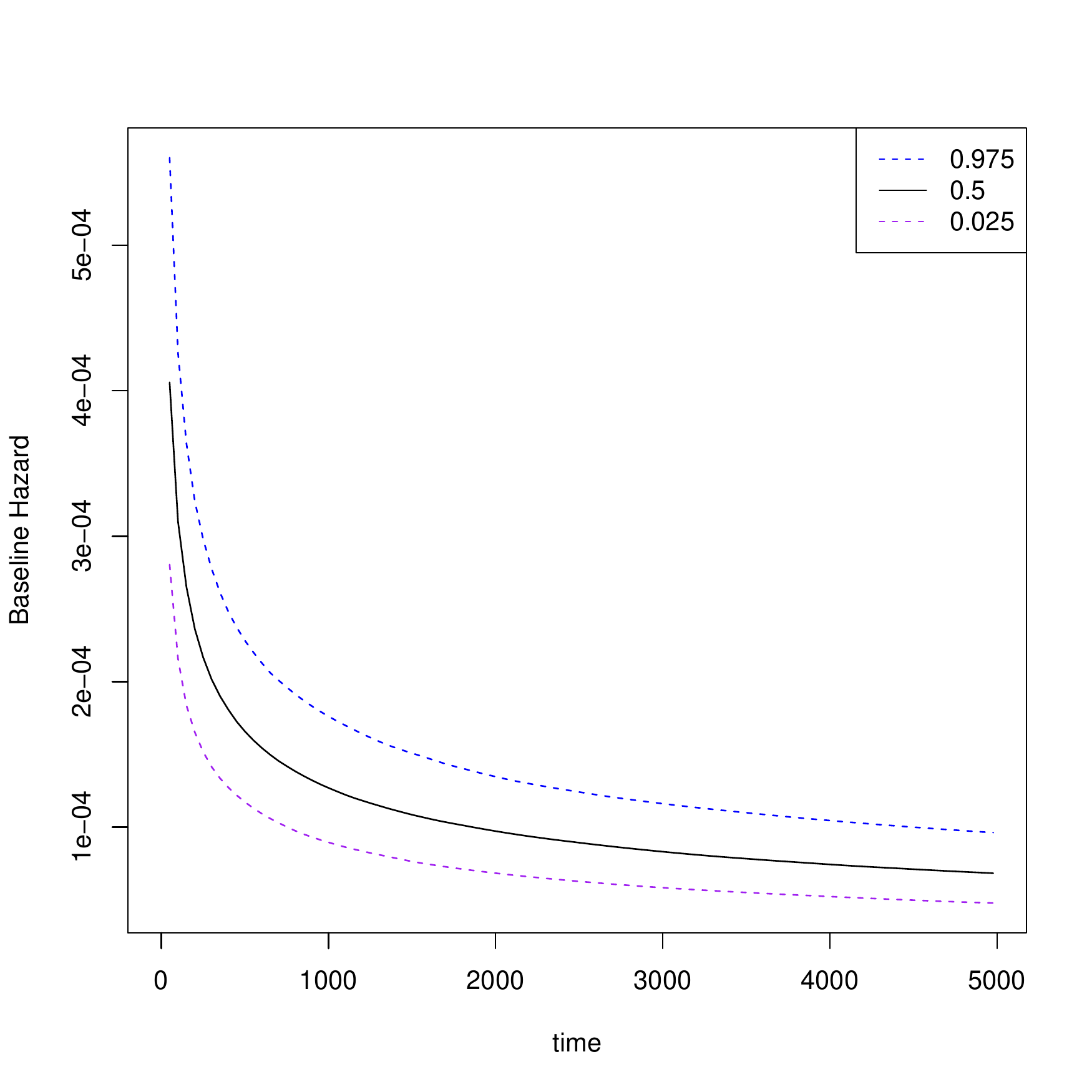}
    \end{minipage}\begin{minipage}{0.5\textwidth}
        \includegraphics[width=0.9\textwidth]{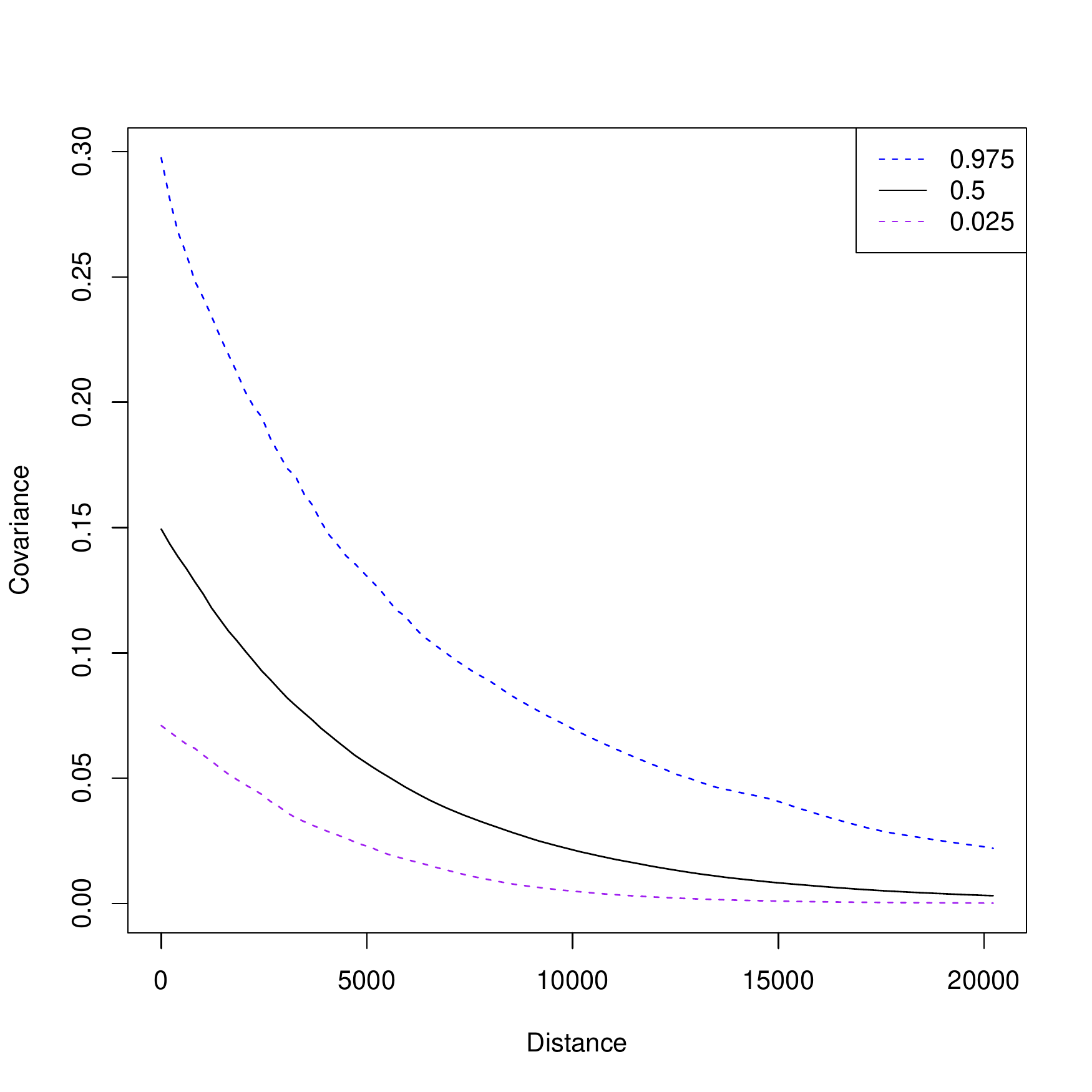}
    \end{minipage}
    
    \caption{\label{fig:leukplots} Top left: predicted $\exp\{Y\}$, to be compared with Figure 6 from \cite{henderson2002}. Top right: plot of $\P[\exp\{Y\}>1.2]$. Bottom left: plot showing the posterior median baseline hazard and 95\% credible interval. Bottom right: plot showing the posterior median of the spatial covariance function and 95\% credible interval.}

\end{figure}

It is also of interest to compare a plot of the mean spatially-correlated frailties $\exp\{Y\}$ with that presented in Figure 6 of \cite{henderson2002}, the former are shown in the top-left plot of Figure \ref{fig:leukplots} -- and comparing this to the original, it can be seen that there is excellent corroboration.

Since we have at our disposal a sample from the joint posterior of all model parameters, we are at liberty to construct other summary measures of interest. For example, the plot of the mean surface $\exp\{Y\}$ is in some sense deficient, as it gives no indication as to how precise the estimates of the latent field are at each point. To address this, we can plot exceedance probabilities: the posterior probability that the exponential of the latent field exceeds some threshold $c$. The interpretation of $\exp\{Y\}>c$ for some threshold $c>1$ is as a multiplicative increase in hazard rate at the spatial location of the particular $Y$ concerned. The threshold $c$ should really be set by specialist clinicians. The top right plot of Figure \ref{fig:leukplots} is a plot of exceedance probabilities derived from the new model, in this case showing $\P[\exp\{Y\}>1.2]$. We examined further plots of exceedances over 1.5 and 2 times the hazard rate, but the predicted probabilities were low in these cases. Lastly (but not exhaustively), we can produce plots of the posterior baseline hazard function and spatial covariance function with credible intervals; these are shown in respectively the bottom left and bottom right plots in Figure \ref{fig:leukplots}.

The purpose of this example has been to demonstrate that the new model and methods give comparable results to a previously published study in the area of spatial survival analysis. A direct comparison was not possible as the two models were fundamentally different, but the main conclusions from these analyses would be very similar.

\section{Discussion}
\label{sect:discussion}

In this article, we have shown how Fourier methods can be used to improve computational performance in spatial survival analysis and  how the proposed method simultaneously solves the problem of spatial prediction. 

Most importantly, the new method can be used for Monte Carlo inference with large datasets for which it would be prohibitively slow to implement the standard $O(n^3)$ method, thus making MCMC tolerable for this class of models. The speedup to $O(n)$ for fixed output grid size represents a substantial gain in computational time and the $O(m\log m)$ increase in cost for increasing grid-size represents (to the author's knowledge) a currently optimal complexity in this respect. The fact that we have provided open-source software for implementing the proposed method \citep{taylor2014a} means that researchers interested in fitting the models discussed in this article on their own datasets will be able to do so with minimal programming effort. 

One issue not discussed in the main body of this article is that the proposed method may not be suitable for learning about very fine scale spatial variation. The limit will in practice be governed by the grid size used, which itself may be dictated by the availability of computational resources. If very fine scale inference is required, one option would be to run the analysis on a smaller observation window of interest, but there must be sufficiently many observations in the smaller window to make this worthwhile. As noted in the discussion of \cite{banerjee2008}: ``learning about fine scale spatial dependence is always a challenge".


This research arose out of the desire to solve a particular problem in survival analysis, but the idea is more widely applicable as will now be illustrated. More generally, the proposed method can be used for any type of classical geostatistical analyses \citep{diggle1998,hoss2010} in which data have been recorded at point-locations and it is desired to (i) make inferences about the effects of individual-level covariates, whilst (ii) adjusting for potential spatial-correlation between observations through some unobserved environmental exposure and (iii) make predictions of functions of the latent field at points in space where we do not have data. This includes semi-parametric and proportional odds spatial survival models, which can be thought of as belonging to the class of geostatistical models. As an example, consider a Poisson geostatistical model for an observed count $Z_i$:
\begin{eqnarray*}
    Z_i &\sim& \text{Poisson}(R_i).\\
    \log R_i &=& X_i\beta + Y_i + U_i,
\end{eqnarray*}
where $Y_i$ is a spatially correlated term and $U_i$ are non-spatial effects. The second of these equations could simply be replaced by
\begin{eqnarray*}
    \log R_i = X_i\beta + Y_\cellof{i} + U_i.
\end{eqnarray*}
whence we again see a computational speed-up from $O(n^3)$ to $O(n)$ for fixed output grid size.

For spatiotemporal datasets, the idea is very similar. Suppose that $\{Y_t\}$ is a stationary spatiotemporal Gaussian process with separable a separable covariance function such that we may write,
\begin{equation*}
    Y_t = a_{t-t,t}Y_{t-1} + [1-a_{t-t,t}]\mu + [1-a_{t-t,t}^2]^{1/2}\epsilon_t,
\end{equation*}
where $a_{t-t,t}\in[0,1]$ and $\epsilon_t\sim\N(0,\Sigma_Y)$. Then if $Y_0\sim\N(\mu,\Sigma_Y)$ this implies that $Y_t\sim\N(\mu,\Sigma_Y)$ for all $t$. Defining the relationship between $\Gamma_t$ and $Y_t$ as $\Gamma_t=\Sigma_Y^{-1/2}(Y_t-\mu)$ or equivalently $Y_t=\mu+\Sigma_Y^{1/2}\Gamma_t$, it can be shown that $\Gamma_t|\Gamma_{t-1}\sim\N[a_{t-t,t}\Gamma_{t-1},1-a_{t-t,t}^2]$, whence assuming $\Gamma_0\sim\N(0,1)$, we are able to evaluate the judicious prior from the decomposition,
\begin{eqnarray*}
    \pi(\Gamma_0,\ldots,\Gamma_T) = \pi(\Gamma_0)\pi(\Gamma_1|\Gamma_0)\cdots\pi(\Gamma_T|\Gamma_{T-1})
\end{eqnarray*}
for all $T$. 

Retaining the same parameter interpretation as above, the spatiotemporal equivalent of the Poisson geostatistical model would then be, 
\begin{eqnarray*}
    Z_{i,t} &\sim& \text{Poisson}(R_{i,t}).\\
    \log R_{i,t} &=& X_{i,t}\beta + Y_\cellof{i,t} + U_{i,t},
\end{eqnarray*}
where for example $Z_{i,t}$ denotes the $i$th observed count at time $t$ and $Y_\cellof{i,t}$ is the value of the process $\{Y_t\}$ at the centroid of the spatiotemporal grid containing this observation. For fixed output grid size, we retain an $O(n)$ increase with the number of observations and $O(m\log m)$ for increasing spatial resolution; increasing the grid resolution in the temporal domain is at a cost of $O(T)$

For users with access to multi-core machines, there is a further MCMC speed-up available, not discussed in the main body of this article. The idea is to use multiple cores to propose new sets of candidate parameters simultaneously, then if the first and second $(\beta^*,\tilde\omega^*,\tilde\eta^*,\Gamma^*)$ are rejected, and the third accepted, say, the MCMC can move using the third proposal. No correction to the acceptance probability is required, because we are not conditioning on the rejected points as in \cite{green2001} for example. This method works because we are targeting a particular acceptance rate, 0.574 in this case, and it is straightforward to compute the efficiency gain in using multiple cores in this way, for the overall acceptance rate assuming each `chance of success' is 0.574 independently on an $k$-core machine follows an upper tail cumulative probability from a $\text{Binomial}(k,0.574)$ probability density function. Thus on an 8-core machine, one can increase the acceptance rate to 99.9\% with an associated relatively small increase in the cost of managing the 8 simultaneous proposals. The relative benefit for MCMC schemes targeting lower acceptance rates, such as 0.234, are even greater; on an 8 core machine, this is upped to 88.1\%. So for the examples in this article, it would be relatively straightforward to reduce the computational time by approximately 40\%, for both the standard and Fourier-based methods.





\section*{Acknowlegements}

The author is very grateful to Professor Robin Henderson for allowing him to access and use an anonymised version of the leukaemia data from \cite{henderson2002}.

\appendix
\section{MCMC Plots}

\begin{figure}[htbp]
    \centering
    \includegraphics[width=0.5\textwidth]{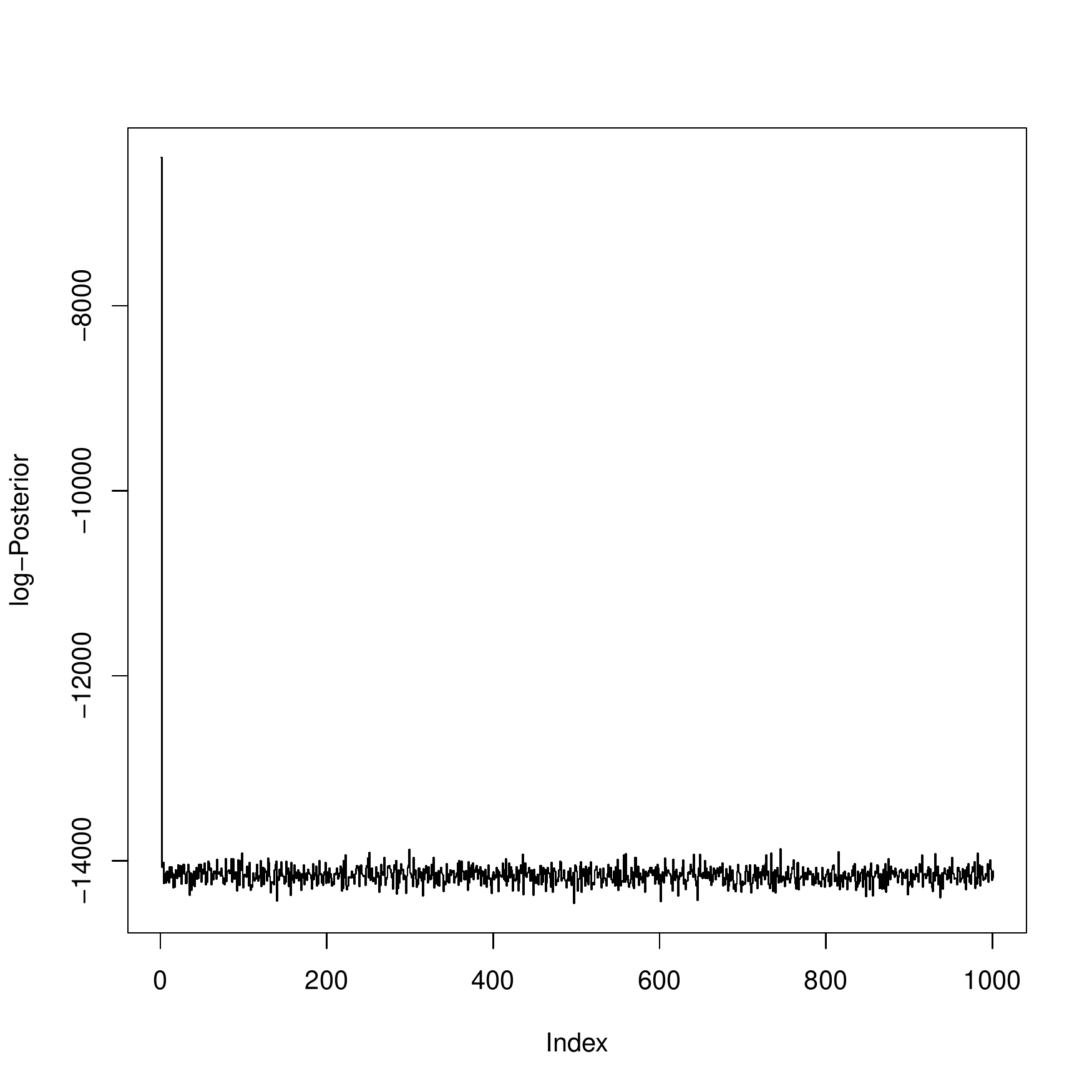}
    \caption{\label{fig:logpost} Plot showing the log of the posterior over the retained iterations. The initial value of the log posterior is also illustrated to show that the chain left the transient phase and appears to have found a mode.}
\end{figure}

\begin{figure}[htbp]
    \centering
    \includegraphics[width=\textwidth]{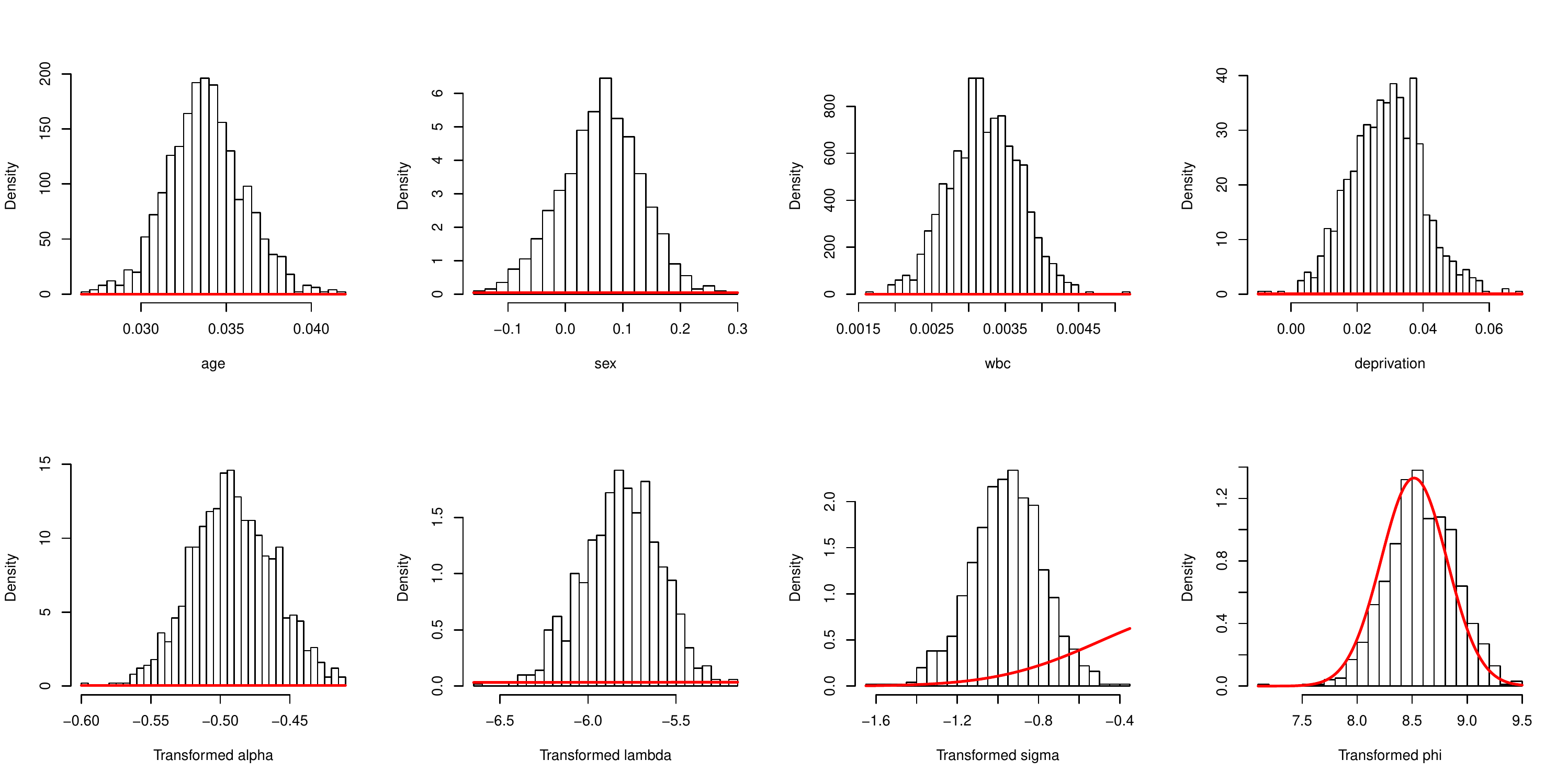}
    \caption{\label{fig:priorposterior} Plot showing the prior (red line) and posterior (histogram) for each parameter.}
\end{figure}

\newpage

\bibliographystyle{chicago}
\bibliography{spatial_survival_bibliography}

\begin{thebibliography}{}

\bibitem[\protect\citeauthoryear{Andrieu and Thoms}{Andrieu and
  Thoms}{2008}]{andrieu2008}
Andrieu, C. and J.~Thoms (2008).
\newblock A tutorial on adaptive {MCMC}.
\newblock {\em Statistics and Computing\/}~{\em 18\/}(4), 343--373.

\bibitem[\protect\citeauthoryear{Banerjee and Carlin}{Banerjee and
  Carlin}{2003}]{banerjee2003a}
Banerjee, S. and B.~P. Carlin (2003).
\newblock Semiparametric spatio-temporal frailty modeling.
\newblock {\em Environmetrics\/}~{\em 14\/}(5), 523--535.

\bibitem[\protect\citeauthoryear{Banerjee and Carlin}{Banerjee and
  Carlin}{2004}]{banerjee2004}
Banerjee, S. and B.~P. Carlin (2004).
\newblock Parametric spatial cure rate models for interval-censored
  time-to-relapse data.
\newblock {\em Biometrics\/}~{\em 60\/}(1), 268--275.

\bibitem[\protect\citeauthoryear{Banerjee and Dey}{Banerjee and
  Dey}{2005}]{banerjee2005}
Banerjee, S. and D.~K. Dey (2005).
\newblock Semiparametric proportional odds models for spatially correlated
  survival data.
\newblock {\em Lifetime data analysis\/}~{\em 11\/}(2), 175--191.

\bibitem[\protect\citeauthoryear{Banerjee, Gelfand, Finley, and Sang}{Banerjee
  et~al.}{2008}]{banerjee2008}
Banerjee, S., A.~E. Gelfand, A.~O. Finley, and H.~Sang (2008).
\newblock Gaussian predictive process models for large spatial data sets.
\newblock {\em Journal of the Royal Statistical Society: Series B (Statistical
  Methodology)\/}~{\em 70\/}(4), 825--848.

\bibitem[\protect\citeauthoryear{Banerjee, Wall, and Carlin}{Banerjee
  et~al.}{2003}]{banerjee2003}
Banerjee, S., M.~M. Wall, and B.~P. Carlin (2003).
\newblock Frailty modeling for spatially correlated survival data, with
  application to infant mortality in {M}innesota.
\newblock {\em Biostatistics\/}~{\em 4\/}(1), 123--142.

\bibitem[\protect\citeauthoryear{Besag and Green}{Besag and
  Green}{1993}]{besag1993}
Besag, J. and P.~J. Green (1993).
\newblock Spatial statistics and {B}ayesian computation.
\newblock {\em Journal of the Royal Statistical Society. Series B
  (Methodological)\/}~{\em 55\/}(1), 25--37.

\bibitem[\protect\citeauthoryear{Bhatt and Tiwari}{Bhatt and
  Tiwari}{2014}]{bhatt2014}
Bhatt, V. and N.~Tiwari (2014).
\newblock A spatial scan statistic for survival data based on {W}eibull
  distribution.
\newblock {\em Statistics in Medicine\/}~{\em 33\/}(11), 1867--1876.

\bibitem[\protect\citeauthoryear{Breslow}{Breslow}{1974}]{breslow1974}
Breslow, N. (1974).
\newblock Covariance analysis of censored survival data.
\newblock {\em Biometrics\/}~{\em 30\/}(1), pp. 89--99.

\bibitem[\protect\citeauthoryear{Brezger, Kneib, and Lang}{Brezger
  et~al.}{2005}]{brezger2005}
Brezger, A., T.~Kneib, and S.~Lang (2005, 9).
\newblock Bayesx: Analyzing bayesian structural additive regression models.
\newblock {\em Journal of Statistical Software\/}~{\em 14\/}(11), 1--22.

\bibitem[\protect\citeauthoryear{Brix and Diggle}{Brix and
  Diggle}{2001}]{brix2001}
Brix, A. and P.~J. Diggle (2001).
\newblock Spatiotemporal prediction for log-{G}aussian {C}ox processes.
\newblock {\em {Journal of the Royal Statistical Society, Series B}\/}~{\em
  63\/}(4), 823--841.

\bibitem[\protect\citeauthoryear{Cox and Oakes}{Cox and Oakes}{1984}]{cox1984}
Cox, D. and D.~Oakes (1984).
\newblock {\em Analysis of Survival Data}.
\newblock Chapman \& Hall/CRC Monographs on Statistics \& Applied Probability.
  Taylor \& Francis.

\bibitem[\protect\citeauthoryear{Cressie and Johannesson}{Cressie and
  Johannesson}{2008}]{cressie2008}
Cressie, N. and G.~Johannesson (2008).
\newblock Fixed rank kriging for very large spatial data sets.
\newblock {\em Journal of the Royal Statistical Society: Series B (Statistical
  Methodology)\/}~{\em 70\/}(1), 209--226.

\bibitem[\protect\citeauthoryear{Damien, Wakefield, and Walker}{Damien
  et~al.}{1999}]{damien1999}
Damien, P., J.~Wakefield, and S.~Walker (1999).
\newblock {G}ibbs sampling for {B}ayesian non-conjugate and hierarchical models
  by using auxiliary variables.
\newblock {\em Journal of the Royal Statistical Society: Series B (Statistical
  Methodology)\/}~{\em 61\/}(2), 331--344.

\bibitem[\protect\citeauthoryear{Darmofal}{Darmofal}{2009}]{darmofal2009}
Darmofal, D. (2009).
\newblock {B}ayesian spatial survival models for political event processes.
\newblock {\em American Journal of Political Science\/}~{\em 53\/}(1),
  241--257.

\bibitem[\protect\citeauthoryear{{De Boor}}{{De Boor}}{1978}]{deboor1978}
{De Boor}, C. (1978).
\newblock {\em A Practical Guide to Splines}.
\newblock Applied Mathematical Sciences. Springer-Verlag.

\bibitem[\protect\citeauthoryear{Diggle, Moyeed, Rowlingson, and
  Thomson}{Diggle et~al.}{2002}]{diggle2002}
Diggle, P., R.~Moyeed, B.~Rowlingson, and M.~Thomson (2002).
\newblock Childhood malaria in the {G}ambia: a case-study in model-based
  geostatistics.
\newblock {\em Journal of the Royal Statistical Society C\/}~{\em 51},
  493--506.

\bibitem[\protect\citeauthoryear{Diggle and Ribeiro~Jr.}{Diggle and
  Ribeiro~Jr.}{2007}]{diggle2007}
Diggle, P. and P.~Ribeiro~Jr. (2007).
\newblock {\em Model-Based Geostatistics}.
\newblock New York: Springer.

\bibitem[\protect\citeauthoryear{Diggle, Tawn, and Moyeed}{Diggle
  et~al.}{1998}]{diggle1998}
Diggle, P.~J., J.~A. Tawn, and R.~A. Moyeed (1998).
\newblock {Model-based geostatistics}.
\newblock {\em Journal of the Royal Statistical Society: Series C (Applied
  Statistics)\/}~{\em 47\/}(3), 299–350.

\bibitem[\protect\citeauthoryear{Diva, Dey, and Banerjee}{Diva
  et~al.}{2008}]{diva2008}
Diva, U., D.~K. Dey, and S.~Banerjee (2008).
\newblock Parametric models for spatially correlated survival data for
  individuals with multiple cancers.
\newblock {\em Stat Med\/}~{\em 27\/}(12), 2127--44.

\bibitem[\protect\citeauthoryear{Edwards and Sokal}{Edwards and
  Sokal}{1988}]{edwards1988}
Edwards, R.~G. and A.~D. Sokal (1988).
\newblock Generalization of the {F}ortuin-{K}asteleyn-{S}wendsen-{W}ang
  representation and {M}onte {C}arlo algorithm.
\newblock {\em Physical Review D\/}~{\em 38}, 2009--2012.

\bibitem[\protect\citeauthoryear{Faucett, Schenker, and Taylor}{Faucett
  et~al.}{2002}]{faucett2002}
Faucett, C.~L., N.~Schenker, and J.~M. Taylor (2002).
\newblock Survival analysis using auxiliary variables via multiple imputation,
  with application to {AIDS} clinical trial data.
\newblock {\em Biometrics\/}~{\em 58\/}(1), 37--47.

\bibitem[\protect\citeauthoryear{Frigo and Johnson}{Frigo and
  Johnson}{2011}]{frigo2011}
Frigo, M. and S.~G. Johnson (2011).
\newblock {FFTW} fastest {F}ourier transform in the west.
\newblock \url{http://www.fftw.org/}.

\bibitem[\protect\citeauthoryear{Fr{\"u}hwirth-Schnatter and
  Fr{\"u}hwirth}{Fr{\"u}hwirth-Schnatter and
  Fr{\"u}hwirth}{2007}]{fruhwirth2007}
Fr{\"u}hwirth-Schnatter, S. and R.~Fr{\"u}hwirth (2007).
\newblock Auxiliary mixture sampling with applications to logistic models.
\newblock {\em Computational Statistics \& Data Analysis\/}~{\em 51\/}(7),
  3509--3528.

\bibitem[\protect\citeauthoryear{Fuentes}{Fuentes}{2007}]{fuentes2007}
Fuentes, M. (2007).
\newblock Approximate likelihood for large irregularly spaced spatial data.
\newblock {\em Journal of the American Statistical Association\/}~{\em
  102\/}(477), 321--331.

\bibitem[\protect\citeauthoryear{Gamerman and Lopes}{Gamerman and
  Lopes}{2006}]{gamermanlopes}
Gamerman, D. and H.~F. Lopes (2006).
\newblock {\em {M}arkov Chain {M}onte {C}arlo: Stochastic Simulation for
  {B}ayesian Inference (2nd ed.)}.
\newblock Chpman \& Hall/CRC.

\bibitem[\protect\citeauthoryear{Ganggang, Liang, and Genton}{Ganggang
  et~al.}{2012}]{xu2015}
Ganggang, X., F.~Liang, and M.~G. Genton (2012).
\newblock A bayesian spatio-temporal geostatistical model with an auxiliary
  lattice for large datasets.
\newblock {\em To appear in Statistica Sinica\/}~{\em 25\/}(1).

\bibitem[\protect\citeauthoryear{Gelfand, Diggle, Fuentes, and Guttorp}{Gelfand
  et~al.}{2010}]{hoss2010}
Gelfand, A., P.~Diggle, M.~Fuentes, and P.~Guttorp (Eds.) (2010).
\newblock {\em Handbook of Spatial Statistics}.
\newblock Chapman and Hall.

\bibitem[\protect\citeauthoryear{Gilks, Richardson, and Spiegelhalter}{Gilks
  et~al.}{1995}]{MCMCiP}
Gilks, W., S.~Richardson, and D.~Spiegelhalter (Eds.) (1995).
\newblock {\em {M}arkov Chain {M}onte {C}arlo in Practice}.
\newblock Chapman \& Hall/CRC.

\bibitem[\protect\citeauthoryear{Girolami and Calderhead}{Girolami and
  Calderhead}{2011}]{girolami2011}
Girolami, M. and B.~Calderhead (2011).
\newblock {R}iemann manifold {L}angevin and {H}amiltonian {M}onte {C}arlo
  methods.
\newblock {\em {Journal of the Royal Statistical Society, Series B}\/}~{\em
  73\/}(2), 123--214.

\bibitem[\protect\citeauthoryear{Green and Mira}{Green and
  Mira}{2001}]{green2001}
Green, P.~J. and A.~Mira (2001).
\newblock Delayed rejection in reversible jump {M}etropolis--{H}astings.
\newblock {\em Biometrika\/}~{\em 88}, 1035--1053.

\bibitem[\protect\citeauthoryear{Hastings}{Hastings}{1970}]{hastings1970}
Hastings, W.~K. (1970).
\newblock {M}onte {C}arlo sampling methods using {M}arkov chains and their
  applications.
\newblock {\em Biometrika\/}~{\em 57\/}(1), 97--109.

\bibitem[\protect\citeauthoryear{Henderson, Shimakura, and Gorst}{Henderson
  et~al.}{2002}]{henderson2002}
Henderson, R., S.~Shimakura, and D.~Gorst (2002).
\newblock Modeling spatial variation in leukemia survival data.
\newblock {\em Journal of the American Statistical Association\/}~{\em 97},
  965--972.

\bibitem[\protect\citeauthoryear{Hennerfeind, Brezger, and
  Fahrmeir}{Hennerfeind et~al.}{2006}]{hennerfeind2006}
Hennerfeind, A., A.~Brezger, and L.~Fahrmeir (2006).
\newblock Geoadditive survival models.
\newblock {\em Journal of the American Statistical Association\/}~{\em
  101\/}(475), 1065--1075.

\bibitem[\protect\citeauthoryear{Higdon}{Higdon}{1998}]{higdon1998}
Higdon, D.~M. (1998).
\newblock Auxiliary variable methods for {M}arkov chain {M}onte {C}arlo with
  applications.
\newblock {\em Journal of the American Statistical Association\/}~{\em
  93\/}(442), 585--595.

\bibitem[\protect\citeauthoryear{Holmes, Held, et~al.}{Holmes
  et~al.}{2006}]{holmes2006}
Holmes, C.~C., L.~Held, et~al. (2006).
\newblock {B}ayesian auxiliary variable models for binary and multinomial
  regression.
\newblock {\em Bayesian Analysis\/}~{\em 1\/}(1), 145--168.

\bibitem[\protect\citeauthoryear{Huang, Kulldorff, and Gregorio}{Huang
  et~al.}{2007}]{huang2007}
Huang, L., M.~Kulldorff, and D.~Gregorio (2007).
\newblock A spatial scan statistic for survival data.
\newblock {\em Biometrics\/}~{\em 63\/}(1), 109--118.

\bibitem[\protect\citeauthoryear{Jerrett, Burnett, Ma, Pope~III, Krewski,
  Newbold, Thurston, Shi, Finkelstein, Calle, et~al.}{Jerrett
  et~al.}{2005}]{jerrett2005}
Jerrett, M., R.~T. Burnett, R.~Ma, C.~A. Pope~III, D.~Krewski, K.~B. Newbold,
  G.~Thurston, Y.~Shi, N.~Finkelstein, E.~E. Calle, et~al. (2005).
\newblock Spatial analysis of air pollution and mortality in los angeles.
\newblock {\em Epidemiology\/}~{\em 16\/}(6), 727--736.

\bibitem[\protect\citeauthoryear{Kammann and Wand}{Kammann and
  Wand}{2003}]{kammann2003}
Kammann, E.~E. and M.~P. Wand (2003).
\newblock Geoadditive models.
\newblock {\em Journal of the Royal Statistical Society: Series C (Applied
  Statistics)\/}~{\em 52\/}(1), 1--18.

\bibitem[\protect\citeauthoryear{Klein, Ibrahim, Scheike, van Houwelingen, and
  Van~Houwelingen}{Klein et~al.}{2013}]{klein2013}
Klein, J., J.~Ibrahim, T.~Scheike, J.~van Houwelingen, and H.~Van~Houwelingen
  (2013).
\newblock {\em Handbook of Survival Analysis}.
\newblock Chapman and Hall/CRC Handbooks of Modern Statistical Methods Series.
  Taylor \& Francis Group.

\bibitem[\protect\citeauthoryear{Klein and Moeschberger}{Klein and
  Moeschberger}{2003}]{klein2003}
Klein, J. and M.~Moeschberger (2003).
\newblock {\em Survival Analysis: Techniques for Censored and Truncated Data}.
\newblock Statistics for Biology and Health. Springer.

\bibitem[\protect\citeauthoryear{Krey, Ligges, and Mersmanne}{Krey
  et~al.}{2011}]{krey2011}
Krey, S., U.~Ligges, and O.~Mersmanne (2011).
\newblock {R} package fftw.
\newblock \url{http://cran.r-project.org/web/packages/fftw/index.html}.

\bibitem[\protect\citeauthoryear{Li and Lin}{Li and Lin}{2006}]{li2006}
Li, Y. and X.~Lin (2006).
\newblock Semiparametric normal transformation models for spatially correlated
  survival data.
\newblock {\em Journal of the American Statistical Association\/}~{\em
  101\/}(474).

\bibitem[\protect\citeauthoryear{Li and Ryan}{Li and Ryan}{2002}]{li2002}
Li, Y. and L.~Ryan (2002).
\newblock Modeling spatial survival data using semiparametric frailty models.
\newblock {\em Biometrics\/}~{\em 58\/}(2).

\bibitem[\protect\citeauthoryear{Metropolis, Rosenbluth, Rosenbluth, Teller,
  and Teller}{Metropolis et~al.}{1953}]{metropolis1953}
Metropolis, N., A.~W. Rosenbluth, M.~N. Rosenbluth, A.~H. Teller, and E.~Teller
  (1953).
\newblock Equation of state calculations by fast computing machines.
\newblock {\em The Journal of Chemical Physics\/}~{\em 21\/}(6), 1087--1092.

\bibitem[\protect\citeauthoryear{M{\o}ller, Syversveen, and
  Waagepetersen}{M{\o}ller et~al.}{1998}]{moller1998}
M{\o}ller, J., A.~R. Syversveen, and R.~P. Waagepetersen (1998).
\newblock Log {G}aussian {C}ox processes.
\newblock {\em Scandinavian Journal of Statistics\/}~{\em 25\/}(3), 451--482.

\bibitem[\protect\citeauthoryear{Neal}{Neal}{2011}]{neal2011}
Neal, R.~M. (2011).
\newblock {\em Handbook of Markov Chain Monte Carlo}, Chapter {MCMC} Using
  {Hamiltonian} Dynamics, pp.\  113--162.
\newblock Chapman \& Hall / CRC Press.

\bibitem[\protect\citeauthoryear{Nott and Green}{Nott and
  Green}{2004}]{nott2004}
Nott, D.~J. and P.~J. Green (2004).
\newblock {B}ayesian variable selection and the {S}wendsen-{W}ang algorithm.
\newblock {\em Journal of computational and Graphical Statistics\/}~{\em
  13\/}(1).

\bibitem[\protect\citeauthoryear{Paciorek}{Paciorek}{2007}]{paciorek2007}
Paciorek, C.~J. (2007, 4).
\newblock Bayesian smoothing with gaussian processes using fourier basis
  functions in the spectralgp package.
\newblock {\em Journal of Statistical Software\/}~{\em 19\/}(2), 1--38.

\bibitem[\protect\citeauthoryear{Paik and Ying}{Paik and Ying}{2012}]{paik2012}
Paik, J. and Z.~Ying (2012).
\newblock A composite likelihood approach for spatially correlated survival
  data.
\newblock {\em Computational statistics \& data analysis\/}~{\em 56\/}(1),
  209--216.

\bibitem[\protect\citeauthoryear{Park and Liang}{Park and
  Liang}{2012}]{park2012}
Park, J. and F.~Liang (2012).
\newblock Bayesian analysis of geostatistical models with an auxiliary lattice.
\newblock {\em Journal of Computational and Graphical Statistics\/}~{\em
  21\/}(2), 453--475.

\bibitem[\protect\citeauthoryear{Pitt and Shephard}{Pitt and
  Shephard}{2001}]{pitt2001}
Pitt, M.~K. and N.~Shephard (2001).
\newblock Auxiliary variable based particle filters.
\newblock In {\em Sequential Monte Carlo methods in practice}, pp.\  273--293.
  Springer.

\bibitem[\protect\citeauthoryear{Pitt and Walker}{Pitt and
  Walker}{2005}]{pitt2005}
Pitt, M.~K. and S.~G. Walker (2005).
\newblock Constructing stationary time series models using auxiliary variables
  with applications.
\newblock {\em Journal of the American Statistical Association\/}~{\em
  100\/}(470), 554--564.

\bibitem[\protect\citeauthoryear{Pollock}{Pollock}{2002}]{pollock2002}
Pollock, K.~H. (2002).
\newblock The use of auxiliary variables in capture-recapture modelling: an
  overview.
\newblock {\em Journal of Applied Statistics\/}~{\em 29\/}(1-4), 85--102.

\bibitem[\protect\citeauthoryear{Rao and Teh}{Rao and Teh}{2013}]{rao2013}
Rao, V. and Y.~W. Teh (2013).
\newblock Fast mcmc sampling for {M}arkov jump processes and extensions.
\newblock {\em The Journal of Machine Learning Research\/}~{\em 14\/}(1),
  3295--3320.

\bibitem[\protect\citeauthoryear{Roberts and Rosenthal}{Roberts and
  Rosenthal}{2001}]{roberts2001}
Roberts, G. and J.~Rosenthal (2001).
\newblock Optimal scaling for various {M}etropolis-{H}astings algorithms.
\newblock {\em Statistical Science\/}~{\em 16\/}(4), 351--367.

\bibitem[\protect\citeauthoryear{Rodrigues and Diggle}{Rodrigues and
  Diggle}{2012}]{rodrigues2012}
Rodrigues, A. and P.~Diggle (2012).
\newblock Bayesian estimation and prediction for inhomogeneous spatiotemporal
  log-gaussian cox processes using low-rank models, with application to
  criminal surveillance.
\newblock {\em Journal of the American Statistical Association\/}~{\em
  107\/}(497), 93--101.

\bibitem[\protect\citeauthoryear{Royle and Wikle}{Royle and
  Wikle}{2005}]{royle2005}
Royle, J.~A. and C.~K. Wikle (2005).
\newblock Efficient statistical mapping of avian count data.
\newblock {\em Environmental and Ecological Statistics\/}~{\em 12\/}(2),
  225--243.

\bibitem[\protect\citeauthoryear{Rue and Held}{Rue and
  Held}{2005}]{rueheld2005}
Rue, H. and L.~Held (2005).
\newblock {\em {G}aussian {M}arkov Random Fields}.
\newblock Chapman \& Hall.

\bibitem[\protect\citeauthoryear{Stroud, Stein, and Lysen}{Stroud
  et~al.}{2014}]{stroud2014}
Stroud, J.~R., M.~L. Stein, and S.~Lysen (2014).
\newblock Bayesian and maximum likelihood estimation for {G}aussian processes
  on an incomplete lattice.
\newblock Available from \url{http://arxiv.org/abs/1402.4281}.

\bibitem[\protect\citeauthoryear{Taylor, Davies, Rowlingson, and Diggle}{Taylor
  et~al.}{2013}]{taylor2013}
Taylor, B.~M., T.~M. Davies, B.~S. Rowlingson, and P.~J. Diggle (2013).
\newblock {lgcp}: An {R} package for inference with spatial and spatio-temporal
  log-{G}aussian {C}ox processes.
\newblock {\em Journal of Statistical Software\/}~{\em 52\/}(4), 1--40.

\bibitem[\protect\citeauthoryear{Taylor and Diggle}{Taylor and
  Diggle}{2014}]{taylor2013c}
Taylor, B.~M. and P.~J. Diggle (2014).
\newblock {INLA} or {MCMC}? a tutorial and comparative evaluation for spatial
  prediction in log-{G}aussian {C}ox processes.
\newblock {\em Journal of Statistical Computation and Simulation\/}.
\newblock Pre-print including extended tutorial available from
  \url{http://www.arxiv.org/pdf/1202.1738}.

\bibitem[\protect\citeauthoryear{Taylor and Rowlingson}{Taylor and
  Rowlingson}{2014}]{taylor2014a}
Taylor, B.~M. and B.~S. Rowlingson (2014).
\newblock spatsurv: an {R} package for {B}ayesian inference with spatial
  survival models.
\newblock {\em Submitted\/}.

\bibitem[\protect\citeauthoryear{Tonda, Satoh, Otani, Sato, Maruyama, Kawakami,
  Tashiro, Hoshi, and Ohtaki}{Tonda et~al.}{2012}]{tonda2012}
Tonda, T., K.~Satoh, K.~Otani, Y.~Sato, H.~Maruyama, H.~Kawakami, S.~Tashiro,
  M.~Hoshi, and M.~Ohtaki (2012).
\newblock Investigation on circular asymmetry of geographical distribution in
  cancer mortality of {H}iroshima atomic bomb survivors based on risk maps:
  analysis of spatial survival data.
\newblock {\em Radiation and environmental biophysics\/}~{\em 51\/}(2),
  133--141.

\bibitem[\protect\citeauthoryear{Wikle}{Wikle}{2002}]{wikle2002}
Wikle, C.~K. (2002).
\newblock {\em Spatial Cluster Modelling}, Chapter Spatial modeling of count
  data: A case study in modelling breeding bird survey data on large spatial
  domains, pp.\  199--209.
\newblock Chapman and Hall.

\bibitem[\protect\citeauthoryear{Wikle}{Wikle}{2010}]{wikle2010}
Wikle, C.~K. (2010).
\newblock {\em Handbook of Spatial Statistics. Low Rank Representations for
  Spatial Processes}, Chapter~8, pp.\  107--118.
\newblock Chapman and Hall.

\bibitem[\protect\citeauthoryear{Wood and Chan}{Wood and Chan}{1994}]{wood1994}
Wood, A. T.~A. and G.~Chan (1994).
\newblock Simulation of stationary gaussian processes in $[0,1]^d$.
\newblock {\em Journal of Computational and Graphical Statistics\/}~{\em
  3\/}(4), 409--432.

\bibitem[\protect\citeauthoryear{Zhang}{Zhang}{2004}]{zhang2004}
Zhang, H. (2004).
\newblock Inconsistent estimation and asymptotically equal interpolations in
  model-based geostatistics.
\newblock {\em Journal of the American Statistical Association\/}~{\em
  99\/}(465), 250--261.

\bibitem[\protect\citeauthoryear{Zhao and Hanson}{Zhao and
  Hanson}{2011}]{zhao2011}
Zhao, L. and T.~E. Hanson (2011).
\newblock Spatially dependent polya tree modeling for survival data.
\newblock {\em Biometrics\/}~{\em 67\/}(2), 391--403.

\end{thebibliography}

\end{document}